\documentclass[journal]{IEEEtran}
\usepackage[utf8]{inputenc}
\usepackage{cite,subfig}
\usepackage{amsmath}
\usepackage{mathtools,amssymb, nccmath}
\usepackage{amsfonts}
\usepackage{amsthm}
\usepackage{amsthm}
\usepackage{amsmath,amssymb,amsfonts}
\usepackage{algorithmic}
\usepackage{graphicx}
\usepackage{textcomp}
\usepackage{xcolor}
\usepackage{graphics}
\usepackage{graphicx}
\usepackage{epsfig}
\usepackage{fancyhdr}
\usepackage{wasysym}
\usepackage{latexsym}
\usepackage{subfig}
\usepackage{comment}
\usepackage{multicol}
\usepackage{colortbl}
\usepackage{color}
\usepackage{mathtools}
\usepackage{cite}
\usepackage{tabularx}
\usepackage{array}
\usepackage{relsize}
\usepackage{makecell}
\newtheorem{lemma}{Lemma}
\usepackage[linesnumbered,ruled,vlined]{algorithm2e}
\usepackage[left=2cm,right=2cm,top=2cm,bottom=2cm]{geometry}
\newcommand{\norm}[1]{\left\lVert#1\right\rVert}

\begin{document}

\title{\Large{Asynchronous Distributed Coordinated Hybrid Precoding in Multi-cell mmWave Wireless Networks}}
        
\author{Meesam~Jafri,\textit{ Student Member,~IEEE,} Suraj~Srivastava,\textit{ Member, IEEE,} Sunil Kumar,  Aditya~K.~Jagannatham, \textit{ Senior Member, IEEE,} and Lajos~Hanzo, \textit{ Life Fellow, IEEE}

\thanks{M. Jafri, S. Srivastava, S. Kumar and A. K. Jagannatham are with the Department of Electrical Engineering, Indian Institute of Technology, Kanpur, Kanpur, 208016, India (e-mail: meesam@iitk.ac.in, ssrivast@iitk.ac.in, pskreddy20@iitk.ac.in, adityaj@iitk.ac.in.). L. Hanzo is with the School of Electronics and Computer Science, University of Southampton, Southampton SO17 1BJ, U.K. (e-mail: lh@ecs.soton.ac.uk)}
}

\maketitle
\begin{abstract}
 Asynchronous distributed hybrid beamformers (ADBF) are conceived for minimizing the total transmit power subject to signal-to-interference-plus-noise ratio (SINR) constraints at the users. Our design requires only limited information exchange between the base stations (BSs) of the mmWave multi-cell coordinated (MCC) networks considered. To begin with, a semidefinite relaxation (SDR)-based fully-digital (FD) beamformer is designed for a centralized MCC system. Subsequently, a Bayesian learning (BL) technique is harnessed for decomposing the FD beamformer into its analog and baseband components and construct a hybrid transmit precoder (TPC). However, the centralized TPC design requires global channel state information (CSI), hence it results in a high signaling overhead. An alternating direction based method of multipliers (ADMM) technique is developed for a synchronous distributed beamformer (SDBF) design, which relies only on limited information exchange among the BSs, thus reducing the signaling overheads required by the centralized TPC design procedure. 
 However, the SDBF design is challenging, since it requires the updates from the BSs to be strictly synchronized. As a remedy, an ADBF framework is developed that mitigates the inter-cell interference (ICI) and also control the asynchrony in the system.
 Furthermore, the above ADBF framework is also extended to the robust ADBF (R-ADBF) algorithm that incorporates the CSI uncertainty into the design procedure for minimizing the the worst-case transmit power. Our simulation results illustrate both the enhanced performance and the improved convergence properties of the ADMM-based ADBF and R-ADBF schemes.
\end{abstract}
%---------------------------------------------------------------
%---------------------------------------------------------------
\begin{IEEEkeywords}
mmWave MIMO, multi cell, coordinated beamforming, inter-cell interference, CSI uncertainty
\end{IEEEkeywords}
%---------------------------------------------------------------
\IEEEpeerreviewmaketitle
%---------------------------------------------------------------
\section{Introduction}
Millimeter wave (mmWave) technology offers a significant promise toward achieving the goal of high throughput in next-generation wireless networks \cite{shokri2015millimeter, rappaport2015wideband, alkhateeb2015limited}. This pivotal technology however faces tremendous challenges in its practical realization. This is due to the fact that mmWave signals suffer from high penetration losses and signal blockage, which significantly degrades the received signal strength \cite{yong2006overview, andrews2016modeling, liang2014low, boulogeorgos2021coverage}. Fortunately, the short wavelength of mmWave frequencies enables a large number of antennas to be tightly packed into a compact array, in turn enabling highly directional beamforming which can help to make up for the signal loss that occurs during propagation \cite{guo2023compressed, satyanarayana2018hybrid}. However, conventional transceiver designs rely entirely on baseband (BB) signal processing, which necessitates a separate RF chain for each of the antennas. This leads to insurmountable implementation challenges in practice due to the associated high power requirement, coupled with excessive hardware complexity. This has prompted the development of the hybrid MIMO signal processing transceiver architectures as a viable solution, that employ significantly fewer RF chains than the number of antenna elements, thereby simplifying their practical implementation \cite{pi2011introduction, sohrabi2016hybrid, shafi20175g, singh2023energy}. 

Furthermore, the severe signal blockage in the mmWave regime also leads to both reduced coverage and to signal quality degradation. The deployment of dense mmWave networks having small cells and distributed base stations (BSs) has shown considerable promise toward overcoming these challenges \cite{bai2014coverage, murdock2013consumption}. In such a deployment, coordination among the BSs can substantially enhance the spectral efficiency by reducing the inter-cell interference (ICI) \cite{jafar2004phantomnet, marsch2008multicell}. In a multi-cell coordinated (MCC) network, multiple BSs covering different cells are linked to a control unit (CU) via a fast backhaul network to jointly design the coordinated beamformers. The performance of coordinated beamformers designed for MCC systems depends on the presence of channel state information (CSI) at the BSs. The level of CSI required for the coordinated beamformer design varies depending upon the specific implementation. In general, a centralized design requires global CSI of all the channels of all the BSs and all the users in the system. By contrast, distributed methods only require each BS to have information about its channels to all the users in the MCC network. This is typically referred to as local CSI in the literature. Therefore, distributed beamformer design entails a significantly lower signaling overhead, which is immensely useful in scenarios having a large number of coordinated cells. 

However, a distributed implementation typically requires accurate coordination and synchronization among the BSs, i.e., the information from different BSs required for distributed coordinated beamformer design must be available simultaneously at the CU. Furthermore, the CU has to wait until all the BSs report their respective information. However, in practice, different BSs often have different computation and communication delays. Furthermore, the transmitted updates are also susceptible to packet losses over the backhaul network. In addition, the BSs may also suffer from intermittent failures that can perturb their operation. All these factors may lead to serious delays in computation of the beamformer weights at the CU.
Finally, even when timely updates are available, it is a challenge to acquire accurate CSI between the users and BSs due to practical factors, such as finite-length training sequences used for channel estimation and the quantized feedback of CSI. 
The resultant CSI error degrades the beamformer performance, unless it is incorporated in the beamformer design. Thus, for improved performance and to realize the full benefits of mmWave MCC systems, it is critical to formulate coordinated hybrid beamformer designs that directly take the asynchronicity as well as the CSI uncertainty into consideration. A brief summary of the contributions and of the knowledge-gaps is presented below.
\subsection{Literature review}
\subsubsection*{Fully Digital TPC Designs}
BS coordination in sub-6 GHz MCC systems has been studied in \cite{chrisT, zhang2010cooperative, xiang2012coordinated, otter} which focus on mitigating the inter-cell interference (ICI) and intra-cell interference.
Ng and Huang \cite{chrisT} proposed a cooperative linear transmit precoding (TPC) technique based on soft interference nulling (SIN) for MCC networks under the idealized simplified assumption that both the CSI and data of all the users in the system is available at each of the BSs. However, the framework proposed therein requires the global CSI to be available at each of the BSs and results in a large amount of information exchange. Hence these results characterize the best-case performance.
Zhang \textit{et al.}, in their path-breaking work in \cite{zhang2010cooperative}, conceived signal-to-interference leakage-plus-noise ratio (SILNR) maximization based distributed linear TPC design techniques for cooperative multi-cell systems that rely on reduced information exchange amongst the BSs. 
Xiang \textit{et al.} \cite{xiang2012coordinated} provided coordinated TPC designs for maximizing the signal-to-interference noise ratio (SINR), incorporating realistic individual BS power constraints. He \textit{et al.} \cite{otter} proposed coordinated TPCs for the maximization of the weighted sum energy efficiency in multi-cell MIMO systems.

However, all the treatises reviewed above consider the availability of perfect CSI. But again, due to the aforementioned challenges, acquiring perfect CSI at all the BSs remains an open challenge in practice. Hence, Several researchers have directed their efforts towards designing a robust beamformer by accounting for the CSI uncertainty.
Lakshminarayana \textit{et al.} \cite{lakshminarayana2015coordinated}, proposed a robust distributed TPC design based on the random matrix theory of massive MIMO MCC networks by minimizing the total transmit power subject to the realistic QoS constraints of all the users.
Xie \textit{et al.} \cite{xie2017robust} proposed an interference alignment based robust beamformer for incorporating the CSI uncertainty by minimizing the interference leakage into their power control problem. 
Dhifallah \textit{et al.} \cite{CRAN} proposed robust coordinated distributed beamforming for transmit power minimization, while taking into account realistic practical constraints, such as the QoS, BS power, CSI error, and backhaul capacity.
Chen \textit{et al.} \cite{chen2020multiobjective} proposed a power control algorithm for reducing the average power consumption by also considering the CSI uncertainty in a MCC network. As a further development, the authors of \cite{chen2021robust, chen2022reconfigurable} considered intelligent reflecting surface (IRS)-aided mmWave MIMO networks to improve the performance of the system further. To elaborate, the authors of \cite{chen2021robust} proposed a robust transmission scheme for IRS-aided mobile mmWave networks considering imperfect statistical CSI under random blockages. Their proposed scheme exploited the fact that the angle-of-arrival (AoA)  and angle-of-departure (AoDs) vary slowly in comparison to the complex path gains of mobile channels. Chen \textit{et al.} \cite{chen2022reconfigurable} exploited the dynamic dual-structured sparsity (DDS) of the angular cascaded mmWave MIMO channel of each user equipment (UE) to perform channel tracking, hence significantly reducing the pilot overhead.
\subsubsection*{Analog and Hybrid Beamforming}
A common feature of all the contributions reviewed above is that they consider fully-digital (FD) TPC schemes, which need separate RF chain for each antenna element, hence they are unsuitable for mmWave MIMO systems. Several researchers have therefore developed novel techniques based on analog and hybrid analog-digital beamforming in order to overcome the above challenge.
In \cite{liang2014low}, the authors proposed a low-complexity phased-zero forcing (PZF) hybrid TPC scheme wherein phase-only control is applied in the RF domain, followed by the design of a low-dimensional BB ZF TPC. The authors of \cite{xu2017spectral} presented a  hybrid TPC design for both uplink and downlink scenarios by optimizing both the energy and spectral efficiencies of the system.
Furthermore, Michaloliakos \textit{et al.} \cite{michaloliakos2016joint} presented a cutting-edge for coordinated analog beamformer designed for mmWave MIMO MCC systems by maximizing the sum-rate of all the users considering predefined beam patterns.
Sun \textit{et al.} \cite{sun2018analytical} proposed a SILNR based regularized zero-forcing (ZF) hybrid beamformer (HBF), with the goal of interference mitigation in a mmWave MCC system. 
As a further advance, Castanheira \textit{et al.} \cite{castanheira2017hybrid} obtained the HBF in a distributed scenario, where the RF TPC is applied at the BSs, while the BB TPC is used at the CU for joint transmission. 
Kumar \textit{et al.} \cite{kumar2021blockage} maximized the weighted sum-rate of a coordinated blockage-aware hybrid beamformer by exploiting the successive convex approximation (SCA) framework. 
Bai \textit{et al.} \cite{bai2018cooperative} conceived a novel cooperative multi-user (MU) TPC for improving energy efficiency by exploiting the unique propagation characteristics of the mmWave MIMO channel. 
Zhao \textit{et al.} \cite{zhao2020robust} proposed a robust distributed hybrid TPC for mmWave multi-cell networks by maximizing the sum-rate of the system. The authors employed the penalty dual decomposition (PDD) aided iterative procedure for circumventing the mathematically intractable nature of their beamformer design. However, in their proposed scheme, the CU is required to obtain the CSI estimates of all the users in the system, which incurs an excessive signaling overhead. The authors of \cite{jafri2022robust} developed an ADMM-based synchronous beamformer design for mmWave MCC networks. However, the proposed framework therein requires a high signalling overhead. Furthermore, the authors of \cite{jafri2023cooperative} design a cooperative beamformer for cell-free networks controlled in a centralized fashion, which requires the exchange of both information symbols and global CSI among the BSs. This in turn leads to a high signalling overhead. In \cite{sha2022near}, the authors proposed a near-interference-free (NIF) user scheduling framework that leverages directional beams to avoid any potential interference for mmWave multi-cell networks by harnessing a hybrid architecture relying on multiple RF chains at the BSs. As a further development, the authors of \cite{sha2023versatile} proposed a generalized NIF-based user scheduling, beam scheduling, and power allocation framework by considering multiple objective functions, i.e., sum-rate maximization, minimum user rate maximization, and total transmit power minimization.
\begin{table}[t]
\caption{List of notations}\label{ListOfNot}
\centering
\begin{tabular}{|l|l|}
\hline
$N$ & Number of cells \\ \hline
$N_t$ & Number of transmit antennas at each BS \\ \hline
$N_{\mathrm{RF},n}$ & Number of RF chains at $n$th BS \\ \hline
$K$ & Number of users in each cell \\ \hline
$s_{nk}$ & Information symbol for $\mathrm{UE}_{nk}$ \\ \hline
$\mathbf{G}_{\mathrm{RF},n}$ & RF TPC matrix at $\mathrm{BS}_n$ \\ \hline
$\mathbf{g}_{\mathrm{BB},nk}$ & baseband TPC for $\mathrm{UE}_{nk}$ \\ \hline
$\mathbf{h}_{mnk}$ & Channel vector between $\mathrm{BS}_m$ and $\mathrm{UE}_{nk}$ \\ \hline
$\Gamma_{nk}$ & SINR of $\mathrm{UE}_{nk}$  \\ \hline
$\gamma_{nk}$ & Target SINR of $\mathrm{UE}_{nk}$ \\ \hline
$G$ & Size of dictionary matrix \\ \hline
$\mathbf{v}$ & Global ICI variable \\ \hline
$\mathbf{v}_n$ & Local ICI variable of $\mathrm{BS}_n$  \\ \hline
$\boldsymbol{\nu}_n$ & Dual variable update of $\mathrm{BS}_n$ \\ \hline
$\boldsymbol{\Phi}_T$ & Dictionary matrix \\ \hline
\end{tabular}
\end{table}

However, all the papers reviewed above assume that the BSs are in perfect coordination and synchronization with each other and the updates from the different BSs are available simultaneously at the CU. But again, the different communication and computation delays associated with each BS and the packet losses over the backhaul network lead to serious delays in the computation and action of the beamformers at the CU. Therefore, to address this knowledge-gap in the existing research, this paper proposes a robust asynchronous distributed beamformer (ADBF) for a multi-user MCC mmWave networks, while considering the availability of both perfect and imperfect CSI. 
\begin{table*}[t!]
%\begin{adjustbox}{width = 1 \linewidth, center}
\small\addtolength{\tabcolsep}{-3.6pt}
    \centering
    \caption{CONTRASTING OUR CONTRIBUTIONS TO THE EXISTING LITERATURE} \label{tab:lit_rev}
\begin{tabular}{|l|c|c|c|c|c|c|c|c|c|c|c|c|c|c|c|c|}

    \hline

 & \cite{chrisT} &  \cite{zhang2010cooperative} & \cite{xiang2012coordinated} & \cite{lakshminarayana2015coordinated} & \cite{xie2017robust} & \cite{CRAN} & \cite{chen2020multiobjective} & \cite{michaloliakos2016joint} & \cite{sun2018analytical}  & \cite{castanheira2017hybrid} & \cite{kumar2021blockage} & \cite{bai2018cooperative} & \cite{zhao2020robust} & \cite{jafri2022robust} & \cite{jafri2023cooperative} & Our \\ [0.5ex]

\hline
mmWave communication &  &  &  &  &  &  &  & \checkmark & \checkmark & \checkmark  & \checkmark & \checkmark & \checkmark &  \checkmark & \checkmark & \checkmark\\
\hline
Hybrid architecture &  &  &  &  &  &  &  & \checkmark & \checkmark & \checkmark  & \checkmark & \checkmark & \checkmark &  \checkmark & \checkmark & \checkmark\\
\hline
Multi-cell & \checkmark & \checkmark & \checkmark & \checkmark & \checkmark & \checkmark & \checkmark & \checkmark & \checkmark & \checkmark & \checkmark & \checkmark & \checkmark & \checkmark &  & \checkmark \\
\hline
Multi-user & \checkmark & \checkmark & \checkmark & \checkmark & \checkmark & \checkmark & \checkmark & \checkmark & \checkmark & \checkmark  & \checkmark & \checkmark & \checkmark & \checkmark & \checkmark & \checkmark \\
\hline
%Coordinated TPC & \checkmark & \checkmark & \checkmark & \checkmark & \checkmark & \checkmark & \checkmark & \checkmark & \checkmark & \checkmark & \checkmark &  &  & \checkmark\\
%\hline
Centralized TPC & \checkmark & \checkmark & \checkmark & \checkmark & \checkmark &  & \checkmark & \checkmark & \checkmark & \checkmark  & \checkmark & \checkmark & \checkmark & \checkmark & \checkmark & \checkmark\\
\hline
SDBF &  &  & \checkmark & \checkmark & \checkmark & \checkmark &  &  &  &   &  &  \checkmark & \checkmark & \checkmark &  &  \checkmark\\
\hline
Robust design  &  &  &  & \checkmark  & \checkmark & \checkmark & \checkmark & \checkmark &  &   &  &  & \checkmark & \checkmark &  & \checkmark\\
\hline
ADBF  &   &  &  &  &  &   &  &  &  &  &  &  &  & &  & \checkmark\\
\hline
%Low signaling overhead  &  &  &  &  &  &  &  &  &  &  &  &  &  & \checkmark\\
%\hline
\end{tabular}
%\end{adjustbox}
\end{table*}
\subsection{Contributions}
An asynchronous distributed hybrid beamformer (ADBF) design framework is proposed for minimizing the total transmit power subject to SINR constraints at the users. The proposed novel ADBF design framework is capable of coping with asynchronicity in the system imposed by network delays and BS failures encountered in distributed hybrid beamformer designs. In contrast to the coordinated beamforming of multi-cell mmWave networks \cite{jafri2022robust} and to cooperative beamforming in cell-free networks \cite{jafri2023cooperative} that ignore the synchronicity, the ADBF design also overcomes the problem of staggering that arises when the updates from the different BSs arrive at the CU with different delays.
First, a centralized BL-based hybrid TPC design framework is formulated. Subsequently, an alternating direction method of multipliers (ADMM)-based synchronous distributed beamformer (SDBF) design is developed to mitigate the excess signaling overhead of the centralized design. Finally, we exploit both the centralized and SDBF designs to formulate the ADBF design, while considering both perfect and imperfect CSI. The contributions of this treatise are boldly and explicitly contrasted to the literature in Table \ref{tab:lit_rev} and are summarized below.
\begin{itemize}
\item A model is developed for coordinated hybrid beamforming in a multi-user MCC mmWave network. Then our centralized TPC design problem is formulated by minimizing the transmit power at each BS, while meeting the SINR requirements for each user, which is shown to be non-convex. A novel two-step TPC design technique is proposed for solving the above problem. First, the FD-TPC is obtained via semidefinite relaxation (SDR), which converts the above non-convex problem into a tractable convex optimization problem. Next, a Bayesian learning (BL)-based framework is introduced for decomposing the FD-TPC into its RF and BB constituents \cite{wipf2004sparse}.

\item Subsequently, a SDBF design technique is formulated for reducing the signalling overhead required by the centralized TPC design. For achieving this, an ADMM approach is proposed for our distributed coordinated beamformer design \cite{boyd2004convex}. Again, the main advantage of the distributed TPC design is that each BS only has to exchange local CSI and ICI information, which leads to significant reduction in the signalling overhead. 

\item Next, an ADBF design is proposed for overcoming the challenges of the SDBF design, which requires strict coordination and synchronization among the BSs. The novel ADBF design framework proposed in this paper incorporates asynchronicity in the system to overcome the challenge of network delays and BS failures for distributed hybrid beamformer designs. Moreover, the technique developed has a low signalling overhead, which makes it appealing for practical implementation. Hence, the ADBF design presented is novel in the context of mmWave MCC networks.

\item Our analysis is subsequently extended to include a realistic scenario associated with CSI uncertainty, and the robust ADBF (R-ADBF) is derived that minimizes the transmit power for the worst-case channel uncertainty. In addition to being non-convex, the robust beamformer design optimization problem has infinitely many constraints arising due to the channel induced errors. To overcome this challenge, the S-lemma \cite{boyd2004convex} is invoked for transforming the ADBF optimization problem into having a finite number of constraints, which are convex in nature. This renders the problem mathematically feasible.

 \item Finally, simulation results are presented to illustrate the efficiency of the ADBF design compared to that of the SDBF design and the feasibility of the proposed asynchronous distributed both algorithms with and without CSI uncertainty.
\end{itemize}
\subsection{Organization of the paper}
The rest of the paper is organized as follows. Section  \ref{MainSysModel} introduces the mmWave MIMO MCC system model and mmWave channel models, followed by our SDR-based centralized beamformer design. This is followed by our BL-based hybrid TPC design procedure in Section \ref{BLTPC}. Section \ref{ADMM} describes the ADMM-based ADBF design conceived for mmWave MCC systems. Subsequently, in Section \ref{RBDCSI} our robust ADBF design is derived for scenarios having imperfect CSI. The convergence analysis of the proposed ADBF design is presented in Section \ref{converg}, followed by our simulation results characterizing the effectiveness of the proposed schemes in  Section \ref{sim}  and our conclusions in Section \ref{conclusions}.
\begin{table}[]
\caption{List of acronyms}\label{ListOfAcro}
\centering
\begin{tabular}{|l|l|}
\hline
MCC & Multi-cell coordination \\ \hline
CSI & Channel state information \\ \hline
BS & Base station \\ \hline
ADBF & Asynchronous distributed beamformer \\ \hline
R-ADBF & Robust asynchronous distributed beamformer \\ \hline
SDBF & Synchronous distributed beamformer \\ \hline
SINR & Signal-to-interference-noise ratio \\ \hline
QoS & Quality of service \\ \hline
BL & Bayesian learning \\ \hline
SDR & Semidefinite relaxation \\ \hline
TPC & Transmit precoder \\ \hline
FD & Fully-digital  \\ \hline
RFCs & Radio frequency chains \\ \hline
ICI & Inter-cell interference  \\ \hline
ADMM & Alternating direction method of multipliers \\ \hline
BB & baseband \\ \hline
PSD & Positive semidefinite  \\ \hline
UE & User-equipment \\ \hline
TAs & Transmit antennas \\ \hline
AoD & Angle of departure \\ \hline
\end{tabular}
\end{table}
\subsubsection*{Notations}
The notation used in this paper is described here. Boldfaced lowercase $\mathbf{x}$ and uppercase $\mathbf{X}$ alphabets are used to represent vectors and matrices, respectively. The operators $\mathbb{E}\{\cdot\}$ and $\mathrm{Tr} \left(\mathbf{X}\right)$ denote the expectation operator and trace of the matrix $\mathbf{X}$, respectively. The notations $\mathbf{X} \succeq \mathbf{0}$ and $\mathbf{X} \succ \mathbf{0} $ represent the fact that the matrix $\mathbf{X}$ is positive semidefinite and positive definite , respectively. The function $\mathrm{rank} \left( \mathbf{X}\right)$ denotes the rank of a matrix $\mathbf{X}$ and the quantity $\Vert{\bf x}\Vert$ represents the Euclidean norm of a vector ${\bf x}$. The matrices ${\bf I}$ and ${\bf 0}$ denote the identity matrix and the all-zero matrix/vector of appropriate dimension, respectively. The quantities ${\bf X}^T$ and ${\bf X}^H$ denote the transpose and Hermitian. A brief lists of notations and acronyms are given in Table-\ref{ListOfNot} and \ref{ListOfAcro}, respectively.
\section{MCC mmWave System Model}\label{MainSysModel}
The mmWave multi-cell downlink system consists of $N$ cells, wherein each cell has a single BS. The BS in each cell has $N_t$ transmit antennas (TAs) and $N_{\mathrm{RF},n}$ RF chains (RFCs) obeying $1 \leq N_{\mathrm{RF},n} << N_t$, and serves $K$ single antenna user equipment (UE). We consider a multi-cell coordinated mmWave network in which the BSs are linked to the control unit through a high-capacity backhaul network, as shown in Fig. 1. Our objective is to jointly design the coordinated beamformers for mitigating the intra-cell interference and ICI. 
Let $\mathrm{BS}_n$ and $\mathrm{UE}_{nk}$ denote the $n$th BS and the $k$th UE in the $n$th cell for $k \in \mathcal{K} = \{1,\ldots, K\}$. Furthermore, assume that $s_{nk}$ denotes the information symbol intended for $\mathrm{UE}_{nk}$ of average power unity, i.e., $\mathbb{E}\{\vert s_{{nk}}\vert^{2}\}=1$.
The signal transmitted by $\mathrm{BS}_n$ can be expressed as
\begin{align}\label{eq1}
    &\mathbf{x}_{n}=\sum_{k=1}^{{K}}{\bf G}_{\mathrm{RF},n}{\bf g}_{\mathrm{BB},nk}s_{{nk}}, \ \ \ \ \forall n \in{\cal N}, k \in \mathcal{K},
\end{align}
where $\mathcal{N} = \{1,\ldots, N\}$, and $\mathbf{G}_{\mathrm{RF},n}\in\mathbb{C}^{N_t \times N_{\mathrm{RF},n}}$ and ${\bf g}_{\mathrm{BB},nk} \in\mathbb{C}^{N_{\mathrm{RF},n}\times 1}$ represent the RF and the BB TPCs, respectively, employed by  $\text{BS}_n$ for transmission to $\text{UE}_{nk}$.
The signal $y_{nk} \in \mathbb{C}$ received at the $\text{UE}_{{nk}}$ is splitted into the desired signal, intra-cell and inter-cell interference components in \eqref{eq2b}.
%\begin{eqnarray}
%y_{nk} &=& \overbrace{{{\mathbf h}}_{nnk}^{H}{\bf G}_{\mathrm{RF},n}{\bf g}_{ \mathrm{BB},nk}s_{{nk}}}^\text{desired signal} +
% \overbrace{\sum_{i\neq k}^{{K}}{\bf h}_{{nnk}}^{{H}}{\bf G}_{\mathrm{RF},n}{\bf g}_{\mathrm{BB},ni}s_{{ni}}}^\text{intra-cell interference} \nonumber \\ & + \underbrace{\sum_{m\neq n}^{N}\sum_{i=1}^{{K}}{\bf h}_{{mnk}}^{{H}}{\bf G}_{\mathrm{RF},m}{\bf g}_{\mathrm{BB},mi}s_{{mi}}}_\text{inter-cell interference} + \zeta_{nk}, \nonumber \\ & \forall n \in\, \mathcal{N}, \forall k \in\, \mathcal{K}, \label{eq2b}
%\end{eqnarray}
\begin{figure*}
    \begin{align}
y_{nk} = \overbrace{{{\mathbf h}}_{nnk}^{H}{\bf G}_{\mathrm{RF},n}{\bf g}_{ \mathrm{BB},nk}s_{{nk}}}^\text{desired signal} +
 \overbrace{\sum_{i\neq k}^{{K}}{\bf h}_{{nnk}}^{{H}}{\bf G}_{\mathrm{RF},n}{\bf g}_{\mathrm{BB},ni}s_{{ni}}}^\text{intra-cell interference} + \overbrace{\sum_{m\neq n}^{N}\sum_{i=1}^{{K}}{\bf h}_{{mnk}}^{{H}}{\bf G}_{\mathrm{RF},m}{\bf g}_{\mathrm{BB},mi}s_{{mi}}}^\text{inter-cell interference} + \zeta_{nk}, \forall n, k, \label{eq2b}
\end{align}
\hrule
\end{figure*}
where $\mathbf{h}_{nmk}\in\mathbb{C}^{N_t}$ is the mmWave downlink channel spanning from $\text{BS}_{{n}}$ to $\text{UE}_{{mk}}$ and $\zeta_{nk}$ denotes the zero-mean symmetric additive complex  white Gaussian noise of variance $\sigma_{{nk}}^{2}$. From \eqref{eq2b}, the mathematical expression for the SINR $ \Gamma_{nk}$ of user $\text{UE}_{{nk}}$ is written as
\begin{align}\label{eq3}
\Gamma_{nk} &=
\frac{\left\vert{\bf h}_{nnk}^{H}\mathbf{G}_{\text{RF},n}\mathbf{g}_{\text{BB},nk}\right\vert^{2}}{\begin{matrix}\bigg\{\sum\limits_{i\neq k}^{K}\left\vert{\bf h}_{nnk}^{H}\mathbf{G}_{\text{RF},n}\mathbf{g}_{\text{BB},ni}\right\vert^{2}&\\+\sum\limits_{m\neq n}^{N}\sum\limits_{i=1}^{K}\left\vert{\bf h}_{mnk}^{H}\mathbf{G}_{\text{RF},m}\mathbf{g}_{\text{BB},mi}\right\vert^{2}+\sigma_{nk}^{2}\bigg\}\end{matrix}}.
\end{align}
%The mmWave channel model for this system is described next. 
The propagation environment between the $\text{BS}_m$ and user $\text{UE}_{nk}$ is modeled as a geometric channel having $L$ multipath components \cite{sohrabi2016hybrid}. Under this model, the channel vector $\mathbf{h}_{mnk}\in\mathbb{C}^{N_t}$ between the $\mathrm{BS}_m$ and $\mathrm{UE}_{nk}$ can be expressed as $\mathbf{h}_{mnk}^H = \sqrt{\frac{N_t}{L}}\sum_{l=1}^{L}\alpha_{l,mnk}\mathbf{a}_{T}^H(\theta_{l}^{T})$,
\begin{comment}
    \begin{equation}\label{eq4}
\mathbf{h}_{mnk}^H = \sqrt{\frac{N_t}{L}}\sum_{l=1}^{L}\alpha_{l,mnk}\mathbf{a}_{T}^H(\theta_{l}^{T}),
\end{equation}
\end{comment}
where the quantities $\alpha _{l}$ and $\theta_{l}^{T}\in[0,2\pi]$ denote the channel gain and AoD, respectively, of the $l$th multi-path component and $\mathbf{a}_{T}(\theta_{l}^{T}) \in \mathbb{C}^{N_t \times 1}$ represents the associated array response vector corresponding to the uniformly spaced linear array (ULA), which can be written as $\mathbf{a}_{T}(\theta_{l}^{T})=\frac{1}{\sqrt{N_t}}\left[1,\ e^{j\frac{2\pi}{\lambda_{d}}d\sin(\theta_{l}^{T})},\ \ldots,\ e^{j\frac{2\pi}{\lambda_{d}}(N_t-1)d\sin(\theta_{l}^{T})}\right]^T,$
\begin{comment}
    \begin{align}
    \mathbf{a}_{T}(\theta_{l}^{T})=\frac{1}{\sqrt{N_t}}\left[1,\ e^{j\frac{2\pi}{\lambda_{d}}d\sin(\theta_{l}^{T})},\ \ldots,\ e^{j\frac{2\pi}{\lambda_{d}}(N_t-1)d\sin(\theta_{l}^{T})}\right]^T,
\end{align}
\end{comment}
where $d$ denotes the antenna spacing and $\lambda_{d}$  represents the carrier wavelength. The centralized TPC design procedure is detailed below.
\vspace{-10pt}
\subsection{Centralized beamforming for the MCC system with perfect CSI}\label{PCSI}
To satisfy the QoS requirement of the user $\mathrm{UE}_{nk}$, $\Gamma_{nk}$ must exceed the required target SINR $\gamma_{{nk}}$, i.e., $\Gamma_{nk}\geq\gamma_{{nk}}$. Mathematically, the weighted sum transmit power minimization problem of centralized TPC design, while satisfying the QoS constraints for each user, can be formulated as
\begin{equation}\label{5}
\begin{aligned}
&\min_{{\{\mathbf{G}_{\mathrm{RF},n}\}},\{{\bf g}_{\text{ BB},nk}\}}\sum_{n=1}^{N}\beta_{n}\left(\sum_{k=1}^{{K}}\Vert{\bf G}_{\mathrm{RF},n}{\bf g}_{\mathrm{BB},nk}\Vert^{2}\right)\\
&\displaystyle \;\;\;\;\;\;\;\;\text{s.t.}\;\;\;\;\;\;\;\;\;\Gamma_{nk}\geq\gamma_{{nk}}, \ \ \ \forall {{n}},{k},\\
&\displaystyle \;\;\;\;\;\;\;\;\;\;\;\;\;\;\;\;\;\;\;\;\;\;\; \vert \mathbf{G}_{\mathrm{RF},n}(i,j)\vert = \frac{1}{\sqrt{N_t}}, \ \ \ \  \forall {n},
\end{aligned}
\end{equation}
where $\Gamma_{nk}$ is as defined in \eqref{eq3} and $\beta_{{n}}$ represents the power weighting factor associated with the $n$th BS. The optimization problem in \eqref{5} is non-convex in nature due to the SINR constraints and also owing to the constant magnitude constraints imposed on the entries of each RF TPC. This makes the problem difficult to solve. Therefore, in order to transform it into a tractable problem, a two-step hybrid TPC design procedure is conceived as next. First, the FD optimal TPC ${\bf g}_{{nk}}$ is obtained using the popular SDR technique. The FD-TPC ${\bf g}_{{nk}}$ obtained from this procedure is subsequently decomposed into its corresponding RF and BB components via the BL method. Upon substituting ${\bf g}_{{nk}} = {\bf G}_{\mathrm{RF},n}{\bf g}_{\mathrm{BB},nk}$  into \eqref{5}, the optimization problem can be equivalently reformulated as
\begin{figure}[t!]
\centering
\begin{center}
\includegraphics[scale=0.30]{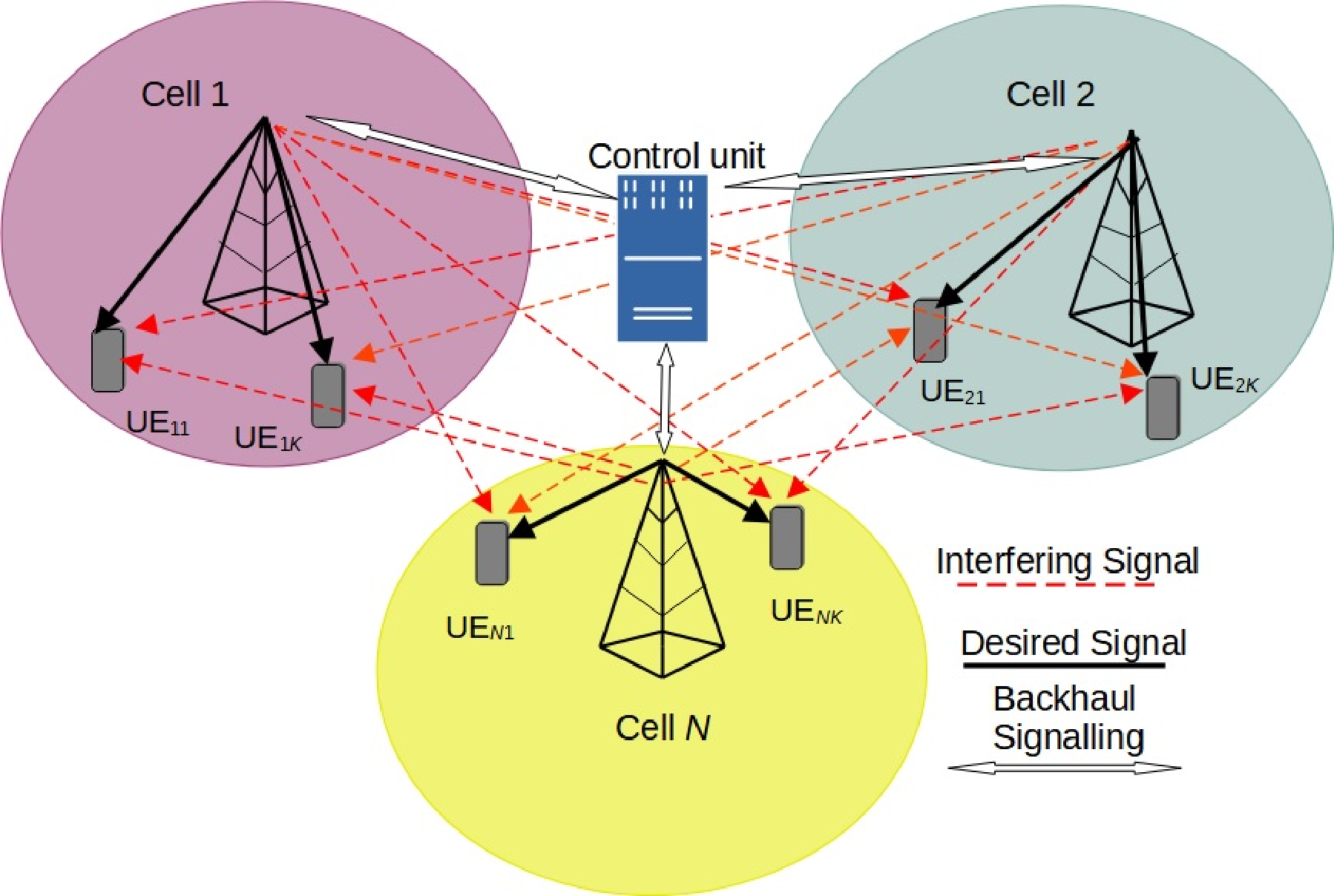}
\end{center}
\caption{Coordinated downlink beamforming in an mmWave MCC network. \vspace{-15pt}}
\label{SystemModel}
\end{figure}
\begin{comment}
    \begin{equation}\label{6}
\begin{aligned}
&\min_{{\{\bf g}_{{nk}}\}} \ \ \sum_{n=1}^{N}\beta_{n}\left(\sum_{k=1}^{{K}}\bf Tr\left({\bf  g}_{nk}{\bf g}_{nk}^H\right) \right)\\
&\displaystyle \;\;\text{s.t.} \;\;\;\;\;\;\;\Gamma_{nk}\geq\gamma_{{nk}}, \ \ \ \forall {{n}},{{k}},
\end{aligned}
\end{equation}
\end{comment}
\begin{align}\label{6}
    \min_{{\{\bf g}_{{nk}}\}} \ \ \sum_{n=1}^{N}\beta_{n}\left(\sum_{k=1}^{{K}}\bf Tr\left({\bf  g}_{nk}{\bf g}_{nk}^H\right) \right)\nonumber \\ \text{s.t.} \;\;\;\; \Gamma_{nk}\geq\gamma_{{nk}}, \ \ \ \forall {{n}},{{k}},
\end{align}
where $\Vert{\bf g}_{nk}\Vert^{2}$ is replaced by $\mathrm{Tr}\left({\bf g}_{nk}{\bf g}_{nk}^H\right)$. The non-convex optimization problem above can be transformed into a convex one via SDR \cite{luo2010sdp}, wherein the matrix ${\bf g}_{{nk}}{\bf g}_{{nk}}^{{H}}$ is replaced by a  rank-1 positive semidefinite (PSD) matrix ${\bf G}_{{nk}}\succeq{\bf 0}$, followed by relaxing the unity rank constraint. The resultant TPC optimization problem can be written as 
\begin{subequations}\label{eq9}
\begin{alignat}{3}
    &\min_{\{{\bf G}_{nk}\}}\sum_{n=1}^{N}\beta_{n}\left(\sum_{k=1}^{{K}}{\rm Tr}({\bf G}_{{nk}})\right)\label{subeq9a}\\& \text{s.t}\quad \frac{1}{\gamma _{{nk}}}\text {Tr}\Big (\mathbf {\bf H}_{{nnk}} \mathbf {\bf G}_{{nk}}\Big)- \sum _{i\neq k}^{{K}} \text{Tr}\Big (\mathbf {\bf H}_{{nnk}} \mathbf {\bf G}_{{ni}}\Big) \geq \nonumber \\
    & \quad\quad\quad\quad \sum_{{m\neq n}}^{N}\sum_{i=1}^{{K}} \text{Tr}\Big (\mathbf {\bf H}_{{mnk}} \mathbf {\bf G}_{{mi}}\Big)+ \sigma_{{nk}} ^{2},\forall {{n}},{{k}}, \label{subeq9b}\\
&{\hskip 21pt}\mathbf{G}_{{nk}} \succeq \mathbf {0},\forall {{n}},{{k}}, \label{subeq9c}
\end{alignat}
\end{subequations}
where \eqref{eq9} is obtained by expanding the quantity $\Gamma_{nk}$ for each $n$, $k$, and ${\bf H}_{mnk}= {\bf h}_{{mnk}}{\bf h}_{mnk}^{H}$. The problem above is a convex semi-definite program (SDP), which can be solved efficiently using widely available tools such as CVX \cite{grant2014cvx}. When the solution $\mathbf{G}_{nk}^{*}$ is of rank-1, the optimal beamformer ${\bf g}_{nk}$  is determined as the eigenvector corresponding to the largest eigenvalue of ${\bf G}_{nk}^{*}$, i.e., ${\bf g}_{nk,\mathrm{opt}}=\sqrt{\lambda}_{\mathrm{max}}{\bf \tilde g}_{nk}$, where ${\bf \tilde g}_{nk}$ denotes the eigenvector having a unit-norm corresponding to $\lambda_{\text{max}}$.
However, when ${\bf G}_{nk}^{*}$ is not a rank-1 matrix, the relaxed problem \eqref{eq9} is not equivalent to the original problem \eqref{6}. Therefore, the solution obtained by solving \eqref{eq9} acts as a lower bound for the problem \eqref{6}. In such cases, the approximate beamformer can be obtained via Gaussian randomization \cite{luo2010sdp}. The state-of-the-art BL method can now be utilized for decomposing the FD-TPC ${\bf g}_{nk,\mathrm{opt}}$ into its constituents RF and BB TPCs, as detailed in the following section.
%-----------------------------------------------------------------
\section{BL-based hybrid precoder design for mmWave MCC systems}\label{BLTPC}
This section presents our hybrid TPC designed for mmWave MCC networks. To begin with, let $\mathbf{G}_{n,\mathrm{opt}} = \left[\mathbf{g}_{n1,\mathrm{opt}}, \mathbf{g}_{n2,\mathrm{opt}}, \ldots, \mathbf{g}_{nK,\mathrm{opt}} \right] \in \mathbb{C}^{N_t \times K}$ denote the concatenated FD beamformer corresponding to all the users at $\mathrm{BS}_n$. The optimization problem of jointly designing the BB and RF TPCs $\mathbf{G}_{\mathrm{BB},n} \in \mathbb{C}^{N_{\mathrm{RF},n} \times K}$ and $\mathbf{G}_{\mathrm{RF},n} \in \mathbb{C}^{N_t \times N_{\mathrm{RF},n}}$, respectively, at $\mathrm{BS}_n$, can be expressed as
\begin{align}
& \left(\mathbf{G}_{\mathrm{BB},n}^{*}, \mathbf{G}_{\mathrm{RF},n}^{*}\right)= \underset{\mathbf{G}_{\mathrm{BB},n}^{*}, \mathbf{G}_{\mathrm{RF},n}^{*}}{\arg \ \ \  \min}\left\|\mathbf{G}_{n,\mathrm{opt}} -\mathbf{G}_{\mathrm{RF},n} \mathbf{G}_{\mathrm{BB},n}\right\|_{\mathrm{F}}\nonumber^{2} \\ 
& \hspace{3cm} \text { s.t. } \ \ \ \ \vert \mathbf{G}_{\mathrm{RF},n}(i,j)\vert = \frac{1}{\sqrt{N_t}}. \label{SBL1}
\end{align}
One can now simplify the above problem as follows. Let $\mathbf{F}_T \triangleq\left[\mathbf{f}_{T}\left(\phi_{1}\right), \mathbf{f}_{T}\left(\phi_{2}\right), \cdots, \mathbf{f}_{T}\left(\phi_{G}\right)\right] \in \mathbb{C}^{N_t \times G}$ denote the dictionary matrix of transmit array response, where $\Phi_T=\{\phi_{g},\ \forall 1\leq g\leq G\}$ represents the quantized set of AoD associated with $\displaystyle \cos(\phi_{g})=\frac{2}{G}(g-1)-1$, and $G$ represents the grid size \cite{heath2016overview}. It is worth noting that the columns of the RF TPC can be appropriately selected from the matrix $\mathbf{F}_T$, since the elements in $\mathbf{F}_T$ meet the constant-magnitude constraint as specified in \eqref{SBL1}. 
In order to achieve the best approximation of the ideal TPC $\mathbf{G}_{n,\mathrm{opt}}$, the equivalent optimization problem of our centralized coordinated hybrid beamformer designed for mmWave MCC networks can be expressed as
\begin{comment}
    \begin{equation}\label{SBLmain}
\begin{array}{c} 
\underset{\tilde{\mathbf{G}}_{\mathrm{BB},n}}{\arg \min}\left\|\mathbf{G}_{n,\mathrm{opt}}-\mathbf{F}_{T} \tilde{\mathbf{G}}_{\mathrm{BB},n}\right\|_{\mathrm{F}}^{2} \\ \hspace{2cm} \text { s.t. } \ \ \ \left\|\operatorname{diag}\left(\tilde{\mathbf{G}}_{\mathrm{BB},n} \tilde{\mathbf{G}}_{\mathrm{BB},n}^{H}\right)\right\|_{0} = N_{\mathrm{RF},n},
\end{array}
\end{equation}
\end{comment}
\begin{align}\label{SBLmain}
& \underset{\tilde{\mathbf{G}}_{\mathrm{BB},n}}{\arg \min}\left\|\mathbf{G}_{n,\mathrm{opt}}-\mathbf{F}_{T} \tilde{\mathbf{G}}_{\mathrm{BB},n}\right\|_{\mathrm{F}}^{2}, \nonumber \\ & \text { s.t. } \ \ \ \left\|\operatorname{diag}\left(\tilde{\mathbf{G}}_{\mathrm{BB},n} \tilde{\mathbf{G}}_{\mathrm{BB},n}^{H}\right)\right\|_{0} = N_{\mathrm{RF},n}, 
\end{align}
where $\tilde{\mathbf{G}}_{\mathrm{BB},n} \in \mathbb{C}^{G \times K}$ represents the intermediate BB TPC. The constraint $\left\|\operatorname{diag}\left(\tilde{\mathbf{G}}_{\mathrm{BB},n} \tilde{\mathbf{G}}_{\mathrm{BB},n}^{H}\right)\right\|_{0} = N_{\mathrm{RF},n}$ is a result of the fact that there are only $N_{\mathrm{RF},n}$ RFCs, implying that the matrix $\tilde{\mathbf{G}}_{\mathrm{BB},n}$ can only have $N_{\mathrm{RF},n}$ non-zero rows. The parameterized Gaussian prior for the matrix $\tilde{\mathbf{G}}_{\mathrm{BB},n}$ can be defined for our BL-based hybrid TPC design as

\begin{align}
p\left(\displaystyle \tilde{\mathbf{G}}_{\mathrm{BB},n};{\boldsymbol{\Gamma}}\right)&= \prod_{i=1}^{G}p\left(\tilde{\mathbf{G}}_{\mathrm{BB},n}(i,:);\gamma_{i}\right) \nonumber \\ &= \prod_{i=1}^{G} \frac{1}{\pi\gamma_i} \exp\left(- \frac{\norm{\tilde{\mathbf{G}}_{\mathrm{BB},n}(i,:)}^2}{\gamma_i}\right), \label{PriorAssignment}
\end{align}
where $\boldsymbol{\Gamma}=\mathrm{diag}\left(\gamma_{1},\ldots,\gamma_{G}\right) \in \mathbb{R}^{G\times G}$  represents the hyperparameter matrix. As observed from \eqref{PriorAssignment}, the hyperparameter $\gamma_i$ is assigned to the $i$th row of the matrix $\tilde{\mathbf{G}}_{\mathrm{BB},n}$, which imposes row sparsity, as seen in the constraint \eqref{SBLmain}. 
The posterior density of the matrix $\tilde{\mathbf{G}}_{\mathrm{BB},n}$ can be expressed as $p\left(\tilde{\mathbf{G}}_{\mathrm{BB},n} \mid \mathbf{G}_{n,\mathrm{opt}};\mathbf{\Gamma}\right) \sim \mathcal{C} \mathcal{N}\left(\boldsymbol{\mathcal{S}}, \mathbf{\Omega}\right)$ in conjunction with
\begin{equation}
\boldsymbol{\mathcal{S}} =\frac{1}{\sigma_{e}^{2}}\mathbf{\Omega}\mathbf{F}_{T}^{H}\mathbf{G}_{\mathrm{opt},n} \ \ \text{and} \ \ \displaystyle \mathbf{\Omega}=\left(\frac{1}{\sigma_{e}^{2}}\mathbf{F}_{T}^{H}\mathbf{F}_{T}+\mathbf{\Gamma}^{-1}\right)^{-1},
\label{eq. mu and sigma}
\end{equation}
 where $\boldsymbol{\mathcal{S}} \in \mathbb{C}^{G \times K}$ and ${\boldsymbol{\Omega}}\in \mathbb{C}^{G \times G}$ represent the \textit{a posteriori} mean matrix and the associated covariance matrix, respectively. Furthermore, $\sigma_{e}^2$ denotes the approximation error variance.  
One can observe that the MMSE estimate, i.e., the \textit{a posteriori} mean $\boldsymbol{\mathcal{S}}$, depends on the hyperparameter matrix $\mathbf{\Gamma}$. Additionally, the $i$th row of the matrix $\tilde{\mathbf{G}}_{\mathrm{BB},n}$, denoted by $\tilde{\mathbf{G}}_{\mathrm{BB},n}(i,:)$, approaches $\mathbf{0}$ as $\gamma_i \rightarrow 0$.
%and this forces the posterior probability to satisfy $p\left(\tilde{\mathbf{F}}_{\mathrm{BB},n}(i) = 0 \mid \mathbf{F}_{n,\mathrm{opt}};\gamma_i = 0\right) = 1$,  which pertains to the $i$th RF precoder column of the dictionary matrix $\mathbf{A}_T$ . 
Therefore, it can be observed that obtaining the estimate of $\tilde{\mathbf{G}}_{\mathrm{BB},n}$ translates into the estimation of the hyperparameter vector $\boldsymbol{\gamma}=[\gamma_{1},\ldots,\gamma_{G}]^{T}$. %Since the pertinent optimization problem toward maximization of the likelihood of $\boldsymbol{\gamma}$ through maximizing the Bayesian evidence $p(\mathbf{G}_{\mathrm{opt}};\boldsymbol{\Gamma)}$ is non-convex.  
The procedure of designing a hybrid TPC using BL can now be utilized for maximizing the Bayesian evidence $p(\mathbf{G}_{\mathrm{opt}};\boldsymbol{\Gamma)}$ by invoking the low-complexity expectation-maximization (EM) method for determining the $\gamma$.

Let us assume that $\hat{\boldsymbol{\Gamma}}^{(j-1)} $ represents the estimate of the hyperparameter matrix $\boldsymbol{\Gamma}$ calculated in the $(j-1)$st iteration. The EM framework has two stages. The expectation stage (E-stage) involves the evaluation of the log-likelihood function $\mathcal{L}\left(\boldsymbol{\Gamma}\mid\hat{\boldsymbol{\Gamma}}^{(j-1)}\right)$ of the hyperparameters, which is given by $\mathbb{E}_{\tilde{{\bf G}}_{\mathrm{BB},n}\mid{\bf G}_{n,\mathrm{opt}};\hat{\boldsymbol{\Gamma}}^{(j-1)}}\Bigg \lbrace\log p\left({\bf G}_{n,\mathrm{opt}},\tilde{{\bf G}}_{\mathrm{BB},n};\boldsymbol{\Gamma}\right)\Bigg \rbrace$.
\begin{comment}
    \begin{align*}
 \mathbb{E}_{\tilde{{\bf G}}_{\mathrm{BB},n}\mid{\bf G}_{n,\mathrm{opt}};\hat{\boldsymbol{\Gamma}}^{(j-1)}}\Bigg \lbrace\log p\left({\bf G}_{n,\mathrm{opt}},\tilde{{\bf G}}_{\mathrm{BB},n};\boldsymbol{\Gamma}\right)\Bigg \rbrace.
\end{align*}
\end{comment}
Next, the average log-likelihood is maximized with respect to the hyperparameter vector $\boldsymbol{\gamma}$ in the maximization stage (M-stage). Hence, the hyperparameter estimates
can be evaluated by obtaining the solution to the following
problem
\begin{align}
\hat{\boldsymbol{\gamma}}^{(j)} = \arg \max _{\boldsymbol{\gamma}} \mathbb{E} \Bigg \lbrace\log p\left(\mathbf{G}_{n,\mathrm{opt}} \mid \tilde{\mathbf{G}}_{\mathrm{BB},n}\right)\nonumber \\ + \log p\left(\tilde{\mathbf{G}}_{\mathrm{BB},n} ; \boldsymbol{\Gamma}\right)\Bigg \rbrace. \label{eqHyp}
\end{align}
In the above equation, it can be observed that the first term inside the $\mathbb{E}\{\cdot\}$ operator is independent of the hyperparameter $\boldsymbol{\gamma}$, and can therefore be omitted in the following M-stage. As a result, the equivalent optimization problem of the M-stage used for determining the hyperparameter estimates can be framed as
\begin{align}\label{estimate_gamma}
\hat{\boldsymbol{\gamma}}^{(j)} =&\, \arg \max _{\boldsymbol{\gamma}} \mathbb{E}_{\tilde{\mathbf{G}}_{\mathrm{BB},n} \mid \mathbf{G}_{n,\mathrm{opt}},\hat{\boldsymbol{\Gamma}}^{(j-1)}}\left\{\log p\left(\tilde{\mathbf{G}}_{\mathrm{BB},n} ; \boldsymbol{\Gamma}\right)\right\}\nonumber \\
%=&\, \arg \max _{\boldsymbol{\gamma}} \sum_{i=1}^{G}\left[ -\log \left( \gamma_{i}\right)-\frac{\mathbb{E}\left(\norm{\tilde{\mathbf{G}}_{\mathrm{BB},n}(i,:)}^{2}_{2}\right)}{\gamma_{i}}\right], \nonumber \\
  =&\, \arg \max _{\boldsymbol{\gamma}} \sum_{i=1}^{G} -\log \left( \gamma_{i}\right)-\frac{\norm{\boldsymbol{\mathcal{S}}^{(j)}(i,:)}^{2} + K \boldsymbol{\Omega}_{(i, i)}^{(j)}}{\gamma_{i}},
\end{align}
where $\boldsymbol{\mathcal{S}}^{(j)}$ and $\boldsymbol{\Omega}^{(j)}$ are obtained from (\ref{eq. mu and sigma}) by setting $\boldsymbol{\Gamma} = \hat{\boldsymbol{\Gamma}}^{(j-1)}$. One can now evaluate the gradient of the objective function (OF) in \eqref{estimate_gamma} with respect to $\boldsymbol{\gamma}$ and set it equal to zero to obtain the optimal value of the hyperparameter estimate $\hat{\gamma}_{i}^{(j)}$. Thus, the estimate of each hyperparameter in the $j$th EM-iteration can be formulated as $\displaystyle \hat{\gamma}_{i}^{(j)}= \frac{1}{K} \norm{\boldsymbol{\mathcal{S}}^{(j)}(:,i)}^2+\boldsymbol{\Omega}^{(j)}_{(i,i)}.$
%\begin{equation*}
%\displaystyle \hat{\gamma}_{i}^{(j)}= \frac{1}{K} \norm{\boldsymbol{\mathcal{S}}^{(j)}(:,i)}^2+\boldsymbol{\Omega}^{(j)}_{(i,i)}.
%\end{equation*}
On convergence, the BB and RF TPCs are obtained in the following manner.
Let $\mathcal{A}$ contain the $N_{\mathrm{RF},n}$ indices of the hyperparameters having the highest magnitude. The BB TPC matrix $\mathbf{G}_{\mathrm{BB},n}^{*}$ corresponding to the $n$th BS can be extracted from $\tilde{\mathbf{G}}_{\mathrm{BB},n}$ as $\mathbf{G}_{\mathrm{BB},n}^{*} = \tilde{\mathbf{G}}_{\mathrm{BB},n}\left(\mathcal{A},:\right)$.
Similarly, the RF TPC $\mathbf{G}_{\mathrm{RF},n}^{*}$ can be chosen from $\mathbf{F}_T$ by columns indexed by the set $\mathcal{A}$ as
$\mathbf{G}_{\mathrm{RF},n}^{*} = \mathbf{F}_{T}\left(:,\mathcal{A} \right)$.
\section{Asynchronous Distributed Coordinated Hybrid Beamformer (ADBF) Design}\label{ADMM}
This section describes the ADBF design procedure of MCC mmWave MIMO networks. As part of this process, the BSs send their information to the CU via backhaul links. Since the BSs in the mmWave regime are typically densely deployed with small cell sizes for mitigating the blockage and propagation losses, BS synchronization is challenging to achieve, especially with an increase in the network size. Furthermore, in practice, transmission delays and packet losses occur frequently, which leads to outdated information at some of the BSs and further aggravates the problem of synchronization in such systems. To address these issues, we develop an ADMM-based ADBF design for MCC mmWave MIMO networks. To begin with, the ADMM technique is reviewed next. This is followed by the algorithms conceived for ADMM-based SDBF and ADBF design.  
\subsection{Overview of ADMM}
ADMM constitutes a state-of-the-art optimization procedure that integrates the concept of dual decomposition with that of the augmented Lagrangian method, which is often employed for solving distributed optimization problems \cite{hestenes1969multiplier}. Hence, the ADMM algorithm typically demonstrates greater numerical stability and faster convergence compared to the conventional dual decomposition method, which results in unbounded sub-problems due to lack of strict convexity \cite{boyd2011distributed, zhang2011unified}.
 To demonstrate the concept of ADMM, consider the following optimization problem that has a separable OF:
\begin{subequations}\label{eq8}
\begin{alignat}{3}
&\min_{{\bf x}\in\mathbb{R}^{n},{\bf z}\in\mathbb{R}^{m}}F({\bf x})+G({\bf z})\label{subeq8a}\\
&\displaystyle \;\;\;\;\;\;\mathrm{s.t.}\;\;\;\;\;\;{\bf A}{\bf x}+{\bf B}{\bf z}={\bf y}, \hspace{0.5cm} {\bf x}\in{\cal S}_{1}, \ {\bf z}\in{\cal S}_{2,}, 
\end{alignat}
\end{subequations}
where $F:\mathbb{R}^{n}\mapsto\mathbb{R}$ and $G:\mathbb{R}^{m}\mapsto\mathbb{R}$ are convex functions, ${\bf A}\in\mathbb{R}^{p \times n}$ , ${\bf B}\in\mathbb{R}^{p \times m}$ and ${\bf y}\in \mathbb{R}^{p}$.
${\cal S}_{1}\subset\mathbb{R}^{n}$ and ${\cal S}_{2}\subset\mathbb{R}^{m}$ are non-empty convex sets. The Lagrangian \cite{boyd2011distributed} for \eqref{eq8} is defined as:
\begin{align}\label{eq997}
{L}_{p}({{\bf x},{\bf z},{\boldsymbol{\xi}}})&=F({\bf x})+G({\bf z})+{\boldsymbol{\xi}}^{T}({\bf A}{\bf x}+{\bf B}{\bf z}-{\bf y})\nonumber \\ & +\frac{{c}}{2}\norm{{\bf A}{\bf x}+{\bf B}{\bf z}-{\bf y}}^2.
\end{align}
The ADMM comprises the following steps in the $i$th iteration

\begin{subequations}\label{eq10}
\begin{alignat}{3}
&{\bf x}^{(i+1)}=\mathop{\arg\min}_{{\bf x}\in{\cal S}_{1}}{L}_{p}({{\bf x},{\bf z}^{(i)},{\boldsymbol{\xi}}^{(i)}}),\label{subeq10a} \\
&\displaystyle{\bf z}^{(i+1)}=\mathop{\arg\min}_{{\bf z}\in{\cal S}_{2}}{L}_{p}({{\bf x}^{(i+1)},{\bf z},{\boldsymbol{\xi}}^{(i)}}),\label{subeq10b}\\
&\displaystyle{\boldsymbol{\xi}}^{(i+1)}=\boldsymbol{\xi}^{(i)}+{c}({\bf A}{\bf x}^{(i+1)}+{\bf B}{\bf z}^{(i+1)}-{\bf y}),\label{subeq10c}
\end{alignat}
\end{subequations}
where ${c}>{0}$ is the penalty parameter. Note that \eqref{subeq10a} and \eqref{subeq10b} are the ${\bf x}$-minimization and ${\bf z}$-minimization steps, respectively, and \eqref{subeq10c} is the dual variable update equation. The dual variable $\boldsymbol{\xi}$ is updated via the subgradient method of \cite{boyd2011distributed} with a step size equal to the penalty parameter ${c}$. The ADMM algorithm alternatively performs one iteration relying on the Gauss-Seidel step \cite{yoon1988lower}, followed by the subgradient update harnessed for improving the convergence speed. The next section describes the ADMM-based distributed framework used in SDBF design for MCC mmWave MIMO networks.
\subsection{Synchronous distributed coordinated hybrid beamformer design}\label{SDMCMU}
In order to apply the steps of the ADMM for our SDBF design, the centralized TPC design problem \eqref{eq9}  of MCC systems is reformulated by introducing the auxiliary variables ${ v_{{mnk}}}=\sum_{i=1}^{{K}} \text{Tr}\Big ({\bf H}_{{mnk}}{\bf G}_{{mi}}\Big)$ and ${V_{{nk}}} \ =\sum_{{m\neq n}}^{N}{ v_{{mnk}}}, \forall m, n,k,$
%\begin{align}
 %   { v_{{mnk}}}=\sum_{i=1}^{{K}} \text{Tr}\Big ({\bf H}_{{mnk}}{\bf G}_{{mi}}\Big), {V_{{nk}}} \ =\sum_{{m\neq n}}^{N}{ v_{{mnk}}}, \forall m, n,k, \label{eq100}
%\end{align}
where ${{{v}}_{{mnk}}}$ denotes the inter-BS interference power emanating from ${\text{BS}}_{{m}}$ upon ${\text{UE}}_{{nk}}$ and ${{{V}}_{{nk}}}$ denotes the total interference power imposed by the nearby BSs upon ${\text{UE}}_{{nk}}$. The optimization problem of our centralized beamformer design in \eqref{eq9} can be reformulated as
\begin{subequations}\label{eq14}
\begin{alignat}{6}
&\min_{\{{\bf g}_{nk}\}}\ \ \sum_{n=1}^{N}\sum_{k=1}^{{K}}{\text{Tr}}({\bf G}_{{nk}}),\ \ \forall n,\label{subeq14a}\\
&\;\;\;\;\;\;\;{\bf G}_{{nk}} \succeq \mathbf {0},\ \ \  \ \ \ \ \forall n, k,\label{subeq14b}\\
&\;\;\;\;\;\;{\frac{1}{\gamma _{{nk}}}} \text{Tr}\Big({\bf H}_{{nnk}}{\bf G}_{{nk}}\Big)- \sum _{j\neq k}^{{K}} \text{Tr}\Big ({\bf H}_{{nnk}}{\bf G}_{{ nj}}\Big) \geq \nonumber\\
& \hspace{3cm} {V_{{nk}}}+\sigma_{{nk}} ^{2}, \forall n, k,\label{subeq14c}  \\
&\;\;\;\;\;\;{ v_{{mnk}}}=\sum_{i=1}^{{K}} \text{Tr}\Big ({\bf H}_{{mnk}}{\bf G}_{{mi}}\Big),\ \ \forall m, n, \label{subeq14d}\\& \;\;\;\;\;\;\; {V_{nk}}=\sum_{{m\neq n}}^{N}{v_{{mnk}}} \geq 0.\label{subeq14e}
\end{alignat}
\end{subequations}
Observe from the SINR constraint \eqref{subeq14c} that each ${\text{UE}}_{{nk}}$ in cell $n$ experiences the sum of the interference power ${V}_{{nk}}$ \eqref{subeq14d} from all the other cells rather than being subjected only to the individual inter-BS interference power ${v_{{mnk}}}$. It is interesting to observe that exchanging the subindices ${{m}}$ and ${{n}}$ has no effect on \eqref{subeq14d}, and does not alter the optimization problem. Therefore, upon interchanging the subscripts ${{m}}$ and ${{n}}$ in \eqref{subeq14d}, the constraints in \eqref{subeq14b} to \eqref{subeq14e} can be divided into ${N}$ independent convex sets  as follows:  

\begin{align}\label{eq15}
& {\mathcal C}_n = \Bigg \lbrace \big (\{\mathbf{G}_{nk}\}_{n,k}, \lbrace V_{{nk}}\rbrace _{{k}}, \lbrace v_{{nmk}}\rbrace _{{m,k}}\big) \big |
 \nonumber \\
 & \hspace{1.5cm} v_{{nmk}} = \sum _{i=1}^{{K}} \text{Tr}(\mathbf
 {H}_{{nmk}}\mathbf{G}_{{ni}}), \ \ \ \ \forall {{m}} \ne {{n}},\ \forall {{k}}, \nonumber \\
& \hspace{1.5cm} \frac{1}{\gamma _{{nk}}}\text{Tr}(\mathbf {H}_{{nnk}}\mathbf {G}_{{nk}})-\sum _{j\ne k}^{{K}}\text{Tr}(\mathbf
 {H}_{{nnk}}\mathbf {G}_{{nj}})\nonumber\\
 & \hspace{3cm} \geq V_{{nk}}+\sigma _{{nk}}^2, \,\forall {{k}}, \nonumber \\
& \hspace{1.5cm} \mathbf {G}_{{nk}} \succeq \mathbf {0}, \,V_{{nk}} \geq 0, \,\forall {{k}} \Bigg \rbrace,\forall n\in \mathcal{N} . 
\end{align}  
We additionally define the following new variables:    
\begin{align}\label{eq16}
{\mathbf{v}} = & \big [[v_{121},\ldots, v_{12K}],\ldots, [v_{N(N-1)1},\ldots,
 v_{N(N-1)K}]\big ]^{T}\nonumber\\
 & \quad \in \mathbb {R}^{N(N-1)K},
\end{align}  
\begin{align}\label{eq17}
\mathbf {v}_{n} =& \big [[V_{n1},\ldots, V_{nK}], [v_{n11},\ldots, v_{n1K}],\nonumber\\
& \ldots, [v_{nN1},\ldots, v_{nNK}]\big
 ]^T \in \mathbb {R}_{+}^{NK}, \,\,\forall n ,
\end{align}  
where ${\mathbf{v}}$ gathers all the global ICI variables and  ${\bf v}_{n}$ collects the ICI variables ${\{V_{nk}\}}_{k=1}^{K}$ and ${\{v_{nmk}\}}_{m,k}$ for $m \in N \backslash \{n\}$. The variable ${\bf v}_{n}$ represents the total interference experienced by ${\mathrm{BS}}_{n}$ along with the total interference experienced by the other cells due to ${\text{BS}}_{n}$. Moreover, we have $\mathbf {v}_{n} =\mathbf {W}_{n} {\mathbf{v}}$, where $\mathbf{W}_{n} \in \lbrace 0, 1\rbrace ^{N{K} \times {N}({N}-1){K}}$ denotes the linear mapping matrix. Hence, \eqref{eq14} can be rewritten as:
\begin{align}\label{eq18}
 & \min_{\{{\bf G}_{{nk}}, {\bf v}_{{n}},{\bf v}\}}\sum_{n=1}^{N}\sum_{k=1}^{{K}}\beta_n{\text{Tr}}({\bf G}_{{nk}})\nonumber \\
& \text{s.t.}\ \ \ \left({\{{\bf G}_{{nk}}\}}_{{k}},{\bf v}_{{n}}\right)\in{\cal C}_{{n}}, \ \ \textrm{and} \ \ {\bf v}_{{n}}={\bf W}_{{n}}{\bf v}, \ \forall n.  
\end{align}
 Upon applying the ADMM technique to \eqref{eq18}, the problem can be recast as
\begin{align}\label{eq19}
    \min_{\{{\bf G}_{nk},{\bf v}_{n}, {\bf v}\}}& \Bigg \{\sum_{n=1}^{N}\sum_{k=1}^{{K}}\beta_{n}{\text{Tr}}\left({\bf G}_{{nk}}\right)+ \frac{c}{2} \sum_{n=1}^{N}\left\Vert{\bf W}_{n}{\bf v}-{\bf v}_{n}\right\Vert^{2} \Bigg\} \nonumber \\
& \!\!\!\!\!\!\! \!\!\!\!\!\!\!\!\!\!\!\! \text{s.t.} \ \left(\{{\bf G}_{nk}\}_k,{\bf v}_{n}\right)\in{\cal C}_{n}, \ \ \textrm{and} \ \ {\bf v}_{{n}}={\bf W}_{{n}}{\bf v}, \ \forall n.
\end{align}
Thus, \eqref{eq19} is equivalent to \eqref{eq18}. The augmented Lagrangian of \eqref{eq19} can be recast as
\begin{align}\label{eq20}
 \min_{{\{{\bf G}_{nk}\}}_{k},\atop{\bf v}_{n},p_{n},n=1,\ldots,M} & \sum_{n=1}^{N}\Bigg\{ \sum_{k=1}^{{K}}\beta_{n}{\text{Tr}}({\bf G}_{{nk}})+{c\over 2}\left\Vert{\bf W}_{n}{\bf v}^{(i)}-{\bf v}_{n}\right\Vert^{2}\nonumber\\
 & -\boldsymbol{\nu}_{n}^{(i)^{T}}{\bf v}_{n} \Bigg\} \nonumber \\
&  \displaystyle{\rm s.t.}\ \ \ \ \left({\big\{{\bf G}_{nk}\big\}}_{k},{\bf v}_{n}\right)\in{\cal C}_{n}, \forall n.     
\end{align}
The original problem can now be decoupled into $N$ independent optimization problems for the design of our synchronous distributed beamformer. For the $n$th cell of the MCC mmWave system, the beamformer design problem can now be expressed as

\begin{align} \label{eq21}
 \left\{{\bf v}_{n}^{(i+1)},{\bf G}_{nk} \right\}&= \arg\min \Bigg\{\sum_{k=1}^{{K}}\beta_{n}{\text{Tr}}({\bf G}_{{nk}})\nonumber\\
 & +{c^{(i)}\over 2}\left\Vert{\bf W}_{n}{\bf v}^{(i)}-{\bf v}_{n}\right\Vert^{2} 
 -{\boldsymbol{\nu}_{n}^{(i)}}^T{\bf v}_{n} \Bigg\},\nonumber \\
& \text{s.t.} \ \ \ \left({\big\{{\bf G}_{nk}\big\}}_{k},{\bf v}_{n}\right)\in{\cal C}_{n}.
\end{align}
The above optimization problem \eqref{eq21} is convex in nature for each cell which can readily be solved by employing widely available tools such as CVX \cite{grant2014cvx}. The update equation for the dual variables $\boldsymbol{\nu}_{n}$ can be written as
\begin{equation}\label{eq22}
\boldsymbol{\nu}_{n}^{(i+1)} = \boldsymbol{\nu}_{n}^{(i)}+c^{(i)}\left({\bf W}_{n}{\bf v}^{(i+1)}-{\bf v}_{n}^{(i+1)}\right), \ \ \ \forall n,
\end{equation}
where the intermediate problem of updating the variable $\mathbf{v}^{(i+1)}$ is given by
\begin{align} \label{eq23}
\bf v^{(i+1)}&=\mathop{\arg\min}_{{\bf v}\in\mathbb{R}^{N(N-1)K}}{c^{(i)}\over 2}\sum_{n=1}^{N}\Vert{\bf v}_{n}{\bf v}-{\bf v}_{n}^{(i+1)}\Vert^{2} \nonumber \\ & + \sum_{n=1}^{N}{\boldsymbol{\nu}_{n}^{(i)}}^T{\bf W}_{n}{\bf v}.
\end{align}
Since the problem \eqref{eq23} is quadratic convex in nature, the closed-form solution of the problem given in \eqref{eq23} can be expressed as
\begin{equation}\label{eq24}
%\begin{alignat}{2} 
{\bf v}^{(i+1)}={\bf W}^{\dagger}\left(\tilde{\bf v}^{(i+1)}-{1\over {c}}\tilde{\boldsymbol{\nu}}^{(i)}\right),
\end{equation}
where   $\tilde{\bf v}^{(i+1)}=\left[\left(\mathbf{v}_{1}^{(i+1)}\right)^{T},\ldots,\left({\bf v}_{N}^{(i+1)}\right)^{T}\right]^{T}$ and $\tilde{\boldsymbol{\nu}}^{(i)}=\left[\left(\boldsymbol{\nu}_{1}^{(i)}\right)^{T},\ldots,\left(\boldsymbol{\nu}_{N}^{(i)}\right)^{T}\right]^{T}$.
The proposed ADMM-based distributed beamformer designed for MCC mmWave networks is synchronous in nature, since the CU only updates the global ICI variable $\mathbf{v}$ upon receiving the updates from all the participating BSs in the system. Hence, the procedure described above is termed SDBF.
Note that, given only the knowledge of the local CSI $\mathbf{h}_{nmk}$, the ADMM steps in \eqref{eq21} can be solved separately at each BS in a distributed fashion by only relying on the local CSI. Next, each BS transmits the updated local information $\{{\bf v}_{n}\}$ and $\boldsymbol{\nu}_{n}^{(i)}$  to the CU. The global ICI variable ${\bf v}$ is evaluated iteratively at the CU using $\{{\bf v}_{n}\}$ from \eqref{eq24}, which is then further employed for updating the quantity $\boldsymbol{\nu}_{n}^{(i+1)}$ in \eqref{eq22} at each BS. The above steps are summarized in Algorithm \ref{SDAD}. 

The SDBF algorithm is based on the idealized simplifying assumption that the  updates from all the BSs are synchronized with respect to each other. In other words, in the SDBF design procedure, the CU must wait for all the BSs to complete the update of $\mathbf{v}_n$ before proceeding further. This can often lead to problems in practice, especially when the BSs have different delays, arising for example due to packet losses, communication delays, etc. In such a scenario, the CU has to wait for the slowest BS to complete its update before proceeding to the next iteration. This allows the system to advance only at the rate of updates gleaned from the slowest BS, which erodes its performance. Additionally, the SDBF design completely halts in the event when no updates are received from some of the participating BSs, which can arise frequently in practice. In order to overcome these impediments, we propose an ADMM-based ADBF design that is robust to BS failures and network delays.   
\begin{algorithm}
{\textbf{Initialization}}: ${i}=0$, $\boldsymbol{\nu}_{n}^{(i)} = \boldsymbol{0}$,${\bf v}^{(i)}=\boldsymbol{0}$ and ${c}>0$\;
 \While{(stopping criterion is not satisfied)}{
obtain the local ICI iterate $\{{\bf v}_{n}^{(i+1)}\}$ and ${\mathbf{G}_{nk}}$ in Eq. \eqref{eq21} \;
Transmit the local ICI $\{{\bf v}_{n}^{(i+1)}\}$ and $\{{\boldsymbol{\nu}}_{n}\}$ to the CU \;
CU updates the public ICI value $\{{\bf v}^{(i+1)}\}$ using Eq. \eqref{eq24} \;
Update the dual variables $\{{\boldsymbol{\nu}}_{n}\}$ using Eq. \eqref{eq22} \; 
   ${i} \leftarrow {i+1}$\;
 }
 \caption{SDBF design for mmWave MCC systems}\label{SDAD}
\end{algorithm}
\subsection{Asynchronous distributed beamformer design}\label{ADADMM}
\begin{figure*}[t!]
\centering
\includegraphics[scale = 0.35]{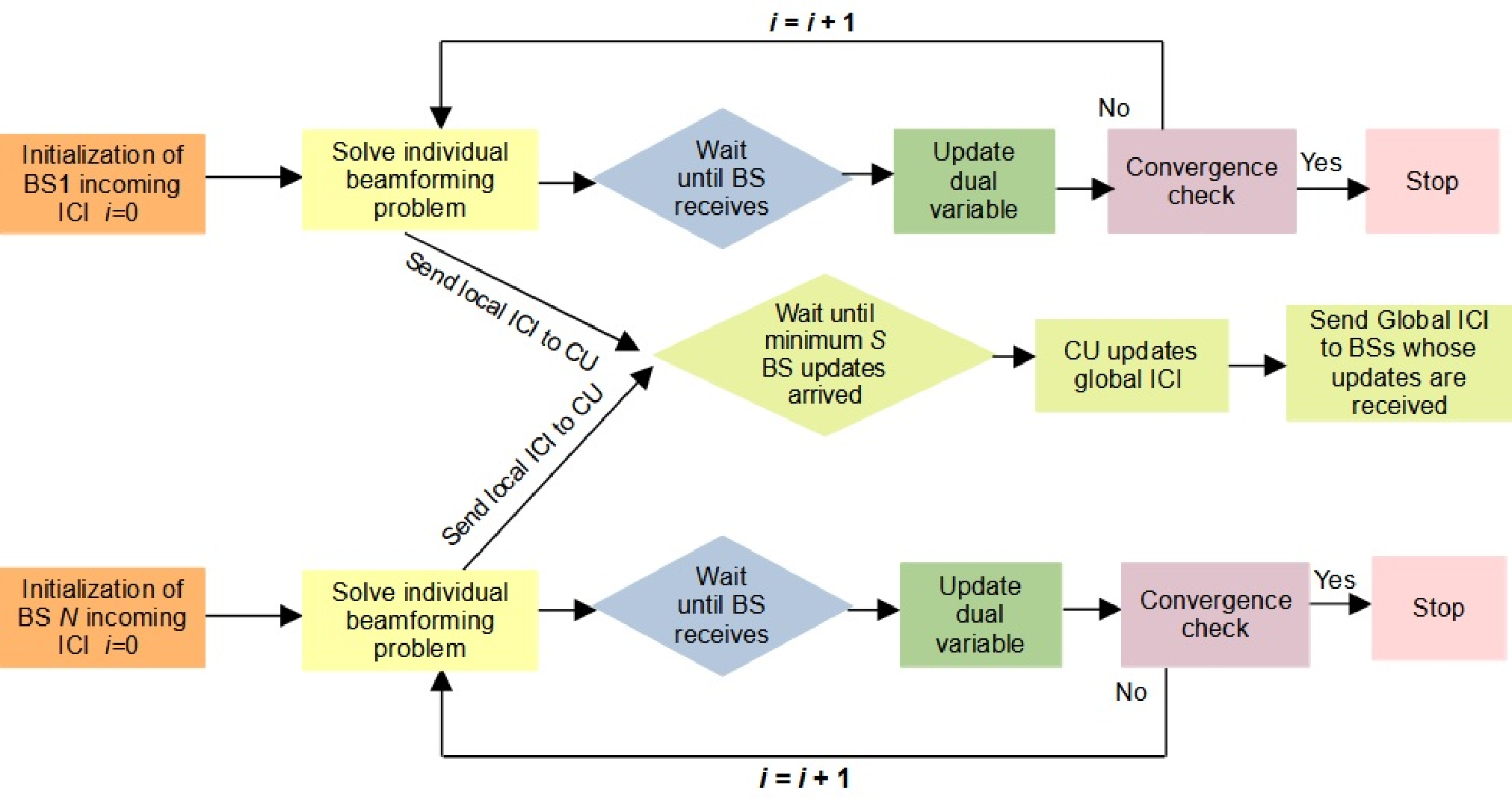}
\caption{Flow-chart of the ADBF design for an $N$-cell scenario \vspace{-15pt}}
\label{ADMMFlowChart}
\end{figure*}
For conceiving an ADBF design, the CU updates the global ICI variable $\mathbf{v}$ upon receiving the local updates $\mathbf{v}_n$ from a subset of the BSs. Hence, neither the BSs have to be synchronized, nor is the CU required to wait for the slowest BS to respond. As a result, under this asynchronous protocol, both the CU and the faster BSs can expedite the variable updates.

As part of this procedure, the CU requires a minimum of $S$ BS responses for updating the global ICI $\mathbf{v}$, where $1 \leq S \leq N$. The ADBF reduces to the SDBF, when $S = N$.
Furthermore, to ensure that all the BSs contribute to the updates and not only the ones that respond the fastest, we also enforce a bounded delay condition. Specifically, the updates from every BS have to be taken into account by the CU at least once in every $\tau$ iterations, where $\tau\geq\ 1$ is a user-defined parameter. A counter $\tau_n$  is maintained by the CU for each $\mathrm{BS}_n$. When the update from BS $n$ arrives at the CU, the corresponding $\tau_n$ is reset to $1$; otherwise, $\tau_n$ is incremented by $1$ as the CU clock $i$ is incremented. 
Let $\mathcal{N}_i \subseteq \mathcal{N}$ denote the index subset of BSs from which the CU receives variable information during iteration $i$.  
Then we have $n \in \mathcal{N}_i$, if the update from $\mathrm{BS}_n$ has arrived at the CU in iteration $i$. For all $n \in \mathcal{N}$ and iteration $i$, it must hold that $n \in \mathcal{N}_i \ {\displaystyle \cup} \ \mathcal{N}_{i-1} \ {\displaystyle \cup} \ldots {\displaystyle \cup} \ \mathcal{N}_{i-\tau+1}$. This implies that an update from BS $n$ must have arrived at least once between iteration $i - \tau + 1$ and iteration $i$ \cite{converg}. Therefore, the variable information used by the CU is at most $\tau$ iterations old. For this assumption to hold, at every iteration, the CU is required to stop and wait for the BSs whose updates have not arrived for $\tau - 1$ iterations, if any. When $\tau_n=\tau$, the CU stops updating the global ICI variable until it receives an update from the $n$th BS, at which time $\tau_n$ is set to $1$. Therefore, the scenario $\tau_n > \tau$ does not arise.
In the ADBF design scheme, the CU also maintains a clock $i$ which starts from zero and it is incremented by $1$ after each update of the variable $\mathbf{v}$. Similarly, each BS has its own independent clock $\{i_n\}_{n=1}^{N}$ that starts from zero and it is incremented by 1 after each dual variable update $\boldsymbol{\nu}_n$.
\subsection{Local ICI update at the BS}
Consider the $n$th BS at time $i_n$. Using the most recent global ICI update $\mathbf{v}$ denoted by $\tilde{\bf v}(n)$ and received by this BS from the CU, the local ICI $\mathbf{v}_n$ is updated via solving 
\begin{align} \label{eq25}
 \left\{{\bf v}_{n}^{(i_n+1)},{\bf G}_{nk}\right\}&= \arg\min \Bigg\{\sum_{k=1}^{{K}}\beta_{n}{\text{Tr}}({\bf G}_{{nk}}) \nonumber \\
& +{c^{(i_n)}\over 2}\left\Vert{\bf W}_{n}{\tilde{\bf v}(n)}-{\bf v}_{n}\right\Vert^{2}-{\boldsymbol{\nu}_{n}^{(i_n)}}^T{\bf v}_{n}\Bigg\},\nonumber \\
& \text{s.t.} \ \ \ \left({\big\{{\bf G}_{nk}\big\}}_{k},{\bf v}_{n}\right)\in{\cal C}_{n},
\end{align}
where the quantity ${\cal C}_{n}$ has been defined in Eq. \eqref{eq15}.
Moreover, since the system is asynchronous in nature, the $\tilde{\bf v}(n)$ are in general different. In other words, some BSs may be using out-of-date versions of the global ICI variable $\mathbf{v}$. The quantities {$\mathbf{v}_n^{(i_n+1)}$ and $\mathbf{\nu}_n^{(i_n)}$} are transmitted next to the CU. Following this, the $n$th BS waits for the next global ICI update $\mathbf{v}$ from the CU before undertaking further processing.
\subsection{Global ICI update by the CU}
When both the bounded delay condition is met and $S$ BS updates arrived, as described in Section \ref{ADADMM}, the CU can proceed with the update of the global ICI variable $\mathbf{v}$. 
%Let $\mathcal{A}_i$ denote the set of BSs whose updates $\mathbf{v}_n$ and $\boldsymbol{\nu}_n$ have arrived at the CU at iteration $i$. 
The global ICI variable $\mathbf{v}$ can be updated as 

\begin{align}\label{eq26}
{\bf v}^{(i+1)}={\bf W}^{\dagger}\left(\tilde{\bf v}^{(i+1)}-{1\over {c}}\tilde{\boldsymbol{\nu}}^{(i)}\right)
\end{align}
where   $\tilde{\bf v}^{(i+1)}=\left[\left(\bf\widehat{v}_{1}\right)^{T},\ldots,\left({\bf \widehat{v}}_{N}\right)^{T}\right]^{T}$ and 
$\tilde{\boldsymbol{\nu}}^{(i)}=\left[\left(\boldsymbol{\widehat{\nu}}_{1}^{(i)}\right)^{T},\ldots,\left(\boldsymbol{\widehat{\nu}}_{N}^{(i)}\right)^{T}\right]^{T}$,
where $\mathbf{\widehat{v}}_n,\mathbf{\widehat{\nu}}_n$ are the most recent updates received from the $n$th BS at the CU. Since as few as $S$ fresh updates might have been received, the update in \eqref{eq26} is still based on $\left\{\mathbf{\widehat{v}}_n,\mathbf{\widehat{\nu}}_n\right\}, \forall 1 \leq n \leq N$. Therefore, it is possible that many of the updates $\left\{\mathbf{\widehat{v}}_n,\mathbf{\widehat{\nu}}_n\right\}$  are outdated.
The CU clock is subsequently incremented by 1, and the updated quantity $\mathbf{v}^{(i+1)}$ is transmitted to only those BSs that have sent updates in the $i$th iteration. Hence, the BSs whose updates are not received by the CU in this iteration will not be aware of this recently updated quantity $\mathbf{v}^{(i+1)}$.  The steps for calculating the global ICI update $\mathbf{v}^{\left(i+1\right)}$ at the CU are given in Algorithm \ref{ADCU}.
\vspace{-10pt}
\subsection{Dual variable update procedure at the BS}
Once the $n$th BS receives the global ICI update ${\bf v}^{\left(i+1\right)}$, it updates the dual variable as follows:
\begin{equation}\label{eq27}
\!\!\boldsymbol{\nu}_{n}^{(i+1)} = \boldsymbol{\nu}_{n}^{(i)}+c^{(i)}\left({\bf W}_{n}{\tilde{\bf v}}-{\bf v}_{n}^{(i_n+1)}\right), \ \ \ \forall n.\\
\end{equation}
The steps required for calculating the local BS updates $\mathbf{v}_n$ at the $n$th BS are given in Algorithm \ref{ADBS}. The BS then increments the local clock $i_n$ by one and the procedure is repeated from step 3 onward in Algorithm \ref{ADBS} until the difference between the $(i-1)$st and $i$th iterations becomes lower than an acceptable tolerance threshold. The flow chart of the ADBF design procedure is displayed in Fig. \ref{ADMMFlowChart} to provide a visual representation. Finally, utilizing the BL-technique developed in Section \ref{MainSysModel}, the corresponding hybrid TPC can once again be derived.
\RestyleAlgo{ruled}
\begin{algorithm}
\caption{Gobal ICI update by the CU}\label{ADCU}
 {\textbf{Initialization}}:{$i=0$,$\widehat{\mathbf{v}}_n = 0$, $\widehat{\boldsymbol{\nu}}_n = 0 $, $n = 1,2,...,N$}
\While{updates received from at least $S$ BSs and $\tau_n\leq \tau, \ \forall n.$}{
\eIf{BS $n$ has transmitted local updates}
{$\tau_n\leftarrow 1$\;
$\mathbf{\widehat{v}}_n\leftarrow\mathbf{v}_n$ (newly received local ICI update from BS $n$)\;
$\mathbf{\widehat{\nu}}_n\leftarrow\mathbf{\nu}_n$ (newly received dual variable update from BS $n$)}{$\tau_n\leftarrow\tau_n+1$\;}

update $\mathbf{v}^{(i+1)}$ using \eqref{eq26}\;
broadcast $\mathbf{v}^{(i+1)}$ to BSs that have sent updates in the current iteration\;
$i\leftarrow i+1$;
}
\end{algorithm}
\RestyleAlgo{ruled}
\begin{algorithm}
\caption{ADBF design for mmWave MCC systems}\label{ADBS}
{\textbf{Initialization}}: {$i_n {=} 0$, $\mathbf{\nu}^{(i_n)}_n = \boldsymbol{0}$,$\mathbf{v}^{(i_n)} = \boldsymbol{0}$ and ${c}>0$}\;
\While{stopping criterion is not satisfied}
{
Obtain the local ICI iterate {$\mathbf{v}^{(i+1)}_n$} using \eqref{eq25}\;
Transmit the local ICI {$\mathbf{v}^{(i+1)}_n$ }and ${\mathbf{\nu}^{(i)}_n}$ to the CU\;
\eIf{(global ICI $\mathbf{v}^{(i+1)}$ update has been received from the CU)}
{
update the dual variables ${\mathbf{\nu}^{(i_n)}_n}$ using \eqref{eq27} respectively\;
}{ wait\;
}
$i_n\leftarrow i_n+1$;
}
\end{algorithm}

\section{Robust Asynchronous distributed Beamformer Design for Imperfect CSI}\label{RBDCSI}
In the previous sections, we developed our ADBF designed for MCC systems considering that the true knowledge of CSI is available at each BS. However, frequently, only imperfect CSI is available as a result of the errors arising due to channel estimation, CSI quantization and feedback, which are inevitable in practical systems. To overcome the effects described above, 
this section extends our framework for robust ADBF designs to consider also realistic scenarios associated with imperfect CSI. To achieve this goal, let the quantities $\hat{\mathbf{h}}_{mnk} \in \mathbb{C}^{N_t}$, $\forall m,n$ denote the estimated CSI modeled as
%\begin{equation}
$\mathbf{h}_{mnk} = \hat{\mathbf{h}}_{mnk} + \boldsymbol{\xi}_{mnk}$. 
%\end{equation}
The quantity $\mathbf{h}_{mnk}$ denotes the true underlying CSI, which is unknown, and the quantity $\boldsymbol{\xi}_{mnk}\in \mathbb{C}^{N_t}$ represents the CSI error that can be modeled as $\boldsymbol{\xi}_{mnk}^{H} \mathbf{R}_{mnk} \boldsymbol{\xi}_{mnk} \leq 1,$
%\begin{equation}\label{errMod}
%\end{equation}
where $\mathbf{R}_{mnk}$ represents a positive definite matrix. This is the popular ellipsoidal channel estimation error model described in \cite{boyd2004convex}. When $\mathbf{R}_{mnk} = \epsilon_{mnk}^{-2}\mathbf{I}_{N_t}$, where $\epsilon_{mnk}^{2} > 0$, the ellipsoidal reduces to the well-known spherical model of CSI uncertainty, formulated as $\norm{\boldsymbol{\xi}_{mnk}}^2 \leq \epsilon_{mnk}^{2}$. 
The robust centralized coordinated TPC design problem that requires the minimum transmit power, while ensuring that the SINR constraint is satisfied for each user for even the most adverse channel in each uncertainty ellipsoid, can be expressed as
 \begin{subequations}\label{eq30}
\begin{alignat}{2}
&\!\!\!\min_{{\{{\bf G}_{\mathrm{RF},n}\}},\{{\bf g}_{\mathrm{BB},nk}\}}\sum_{n=1}^{N}\beta_{n}\left(\sum_{k=1}^{K}\Vert{\bf G}_{\mathrm{RF},n}{\bf g}_{\mathrm{BB},nk}\Vert^{2}\right)\nonumber \\
&\displaystyle \ \text{s.t.}\ \ \Gamma_{nk}\geq\gamma_{nk}, \ \ \forall \ \boldsymbol{\xi}_{mnk}^{H} \mathbf{R}_{mnk} \boldsymbol{\xi}_{mnk} \leq 1, \forall m, n, \label{subeq30a} \\
&\;\;\;\;\;\;\;\;\;\;\;\;\;\;\;\;\;\;\;\;\;\; \vert \mathbf{G}_{\mathrm{RF},n}(i,j)\vert = \frac{1}{\sqrt{N_t}}, \;\;\; \forall n, \label{subeq30b}
\end{alignat}
\end{subequations}
where $\Gamma_{nk}$ is defined in \eqref{iSINR}.
\begin{figure*}
\begin{align}\label{iSINR}
 \Gamma_{nk}= \frac{\left\vert \left(\hat{\mathbf{h}}_{nnk} + \boldsymbol{\xi}_{nnk}\right)^{H} \mathbf{G}_{\mathrm{RF},n}\mathbf{g}_{\mathrm{BB},nk} \right\vert^2}{\sum\limits_{i\neq k}^{K} \left\vert \left(\hat{\mathbf{h}}_{nnk} + \boldsymbol{\xi}_{nnk}\right)^{H}\mathbf{G}_{\mathrm{RF},n}\mathbf{g}_{\mathrm{BB},ni} \right\vert^2+ \sum\limits_{m\neq n}^{N}\sum\limits_{i=1}^{{K}}\left\vert \left(\hat{\mathbf{h}}_{mnk} + \boldsymbol{\xi}_{mnk}\right)^{H}\mathbf{G}_{\mathrm{RF},m}\mathbf{g}_{\mathrm{BB},mi} \right\vert^2 + \sigma^2_{nk}}.
\end{align}
\hrule
\end{figure*}
Note that the constraints \eqref{subeq30b} are non-convex in nature and hence the problem is difficult to solve. To make the problem tractable, similar to Section \ref{PCSI}, we begin with the design of the FD-TPC by setting ${\bf G}_{\mathrm{RF},n}{\bf g}_{\mathrm{BB},nk}= \bf {g}_{nk}$. Using this substitution, the SINR constraint of each UE can be written as
\begin{align}\label{eq32}
&\left(\hat{\mathbf{h}}_{nnk}^{H}\!\!+\!\!\boldsymbol{\xi}_{nnk}^H\right)\!\!\left(\!\!{1\over\gamma_{nk}}{\bf g}_{nk}{\bf g}_{nk}^H\!-\!\sum_{i\neq k}^{K}{\bf g}_{ni}{\bf g}_{ni}^H\!\!\right)\!\!\left(\hat{\mathbf{h}}_{nnk}\!+\!\boldsymbol{\xi}_{nnk}\right)\nonumber \\ & \geq\!\sum_{m\neq n}^{N}\!\!\left(\hat{\mathbf{h}}_{mnk}^{H}\!\!+\!\!\boldsymbol{\xi}_{mnk}^H\right)\!\!\left(\sum_{i=1}^{K}{\bf g}_{mi}{\bf g}_{mi}^H\right)\!(\hat{\mathbf{h}}_{mnk}\!+\!\boldsymbol{\xi}_{mnk}) +\sigma_{nk}^{2}, \nonumber \\ & \ \ \ \ \  \forall \ \boldsymbol{\xi}_{mnk}^{H} \mathbf{R}_{mnk} \boldsymbol{\xi}_{mnk} \leq 1,\forall n, k.
\end{align}
Once again the SDR framework can be employed, wherein ${\bf g}_{nk}{\bf g}_{nk}^H$ is replaced by the PSD matrix ${\bf G}_{nk}\succeq{\bf 0}$ followed by the relaxation of the rank constraint. The above constraint then becomes:
\begin{subequations}\label{eq33}
\begin{alignat}{2}
& \left(\hat{\bf h}_{nnk}^{H}\!\!+\!\!\boldsymbol{\xi}_{nnk}^{H} \right)\left({1\over\gamma_{nk}}{\bf G}_{ nk}\!\!-\!\!\sum_{i\neq k}^{K}{\bf G}_{ni}\right)(\hat{\bf h}_{nnk}\!\!+\!\!\boldsymbol{\xi}_{nnk})\nonumber\\
& \geq\sum_{m\neq n}^{N}(\hat{\bf h}^{H}_{mnK}\!\!+\!\!\boldsymbol{\xi}_{mnk}^{H})\!\!\left(\sum_{i=1}^{K}{\bf G}_{ mi}\right)\!\!\left(\hat{\bf h}_{mnk}\!\!+\!\!\boldsymbol{\xi}_{mnk}\right)+\sigma_{nk}^{2},\nonumber\\
& \ \forall \ \boldsymbol{\xi}_{mnk}^{H} \mathbf{R}_{mnk} \boldsymbol{\xi}_{mnk} \leq 1,\ \ \forall n, k, \ \ \textrm{and} \ \ \mathbf{G}_{nk} \succeq {\bf 0}.
\end{alignat}
\end{subequations}
The SINR constraints of \eqref{eq33} can be rewritten as
\begin{align}\label{eq34}
&\!\!\!\min_{\boldsymbol{\xi}_{nnk}^{H} \mathbf{R}_{nnk} \boldsymbol{\xi}_{nnk} \leq 1}\!\!\left(\!\hat{\bf h}_{nnk}^{H}+\boldsymbol{\xi}_{nnk}^H\!\right)\left(\!\!{1\over\gamma_{nk}}{\bf G}_{nk}-\sum_{i\neq k}^{K}{\bf G}_{ni}\right)\nonumber \\ & \times \left(\hat{\bf h}_{nnk}+\boldsymbol{\xi}_{nnk}\right) \geq\sum_{m\neq n}^{N}\!\Bigg\{\max_{\boldsymbol{\xi}_{mnk}^{H} \mathbf{R}_{mnk} \boldsymbol{\xi}_{mnk} \leq 1}  \left(\!\hat{\bf h}_{mnk}^{H}\!+\!\boldsymbol{\xi}_{mnk}^H \!\right) \nonumber\\
& \hspace{1cm} \times \left(\!\sum_{i=1}^{K}{\bf G}_{mi}\!\right) \left(\hat{\bf h}_{mnk}\!+\!\boldsymbol{\xi}_{mnk}\!\right)\Bigg\}\!+\sigma_{nk}^{2}.
\end{align}
In the above expression, the right-hand term for each $m$ represents the
worst-case ICI power arriving from $\text{BS}_m$ at $\mathrm{UE}_{nk}$, $\forall \ m \in{\cal N} \backslash \{n\}$. Therefore, by defining the slack variable $v_{mnk}$ as

\begin{align}\label{eq35}
v_{mnk}= & \max_{\boldsymbol{\xi}_{mnk}^{H} \mathbf{R}_{mnk} \boldsymbol{\xi}_{mnk} \leq 1}\!\!\!\left(\hat{\bf h}_{mnk}^{H} + \boldsymbol{\xi}_{mnk}^H\right)\!\!\left(\!\sum_{i=1}^{K}{\bf G}_{mi}\!\right)\nonumber\\
&\left(\!\hat{\bf h}_{mnk}\!+\!\boldsymbol{\xi}_{mnk}\!\right),
\end{align}
the constraints can be reformulated as

\begin{subequations}\label{eq36}
\begin{alignat}{3}
& \left(\!\hat{\bf h}_{nnk}^{H} + \boldsymbol{\xi}_{nnk}^H\!\right)\!\left({1\over\gamma_{nk}}{\bf G}_{nk}\!-\!\sum_{i\neq k}^{K}{\bf G}_{ni}\right)\!(\hat{\bf h}_{nnk}+\boldsymbol{\xi}_{nnk})\nonumber\\
& \hspace{-0.5cm} \geq\sum_{m\neq n}^{M}{ v_{mnk}}+\sigma_{nk}^{2},\ \forall\;\;\boldsymbol{\xi}_{nnk}^{H} \mathbf{R}_{nnk} \boldsymbol{\xi}_{nnk} \leq 1, \forall n,k, \label{subeq36b}\\
& \left(\hat{\bf h}_{mnk}^{H}+\boldsymbol{\xi}_{mnk}^{H}\right)\left(\sum_{i=1}^{K}{\bf G}_{mi}\right)\left(\hat{\bf h}_{mnk} + \boldsymbol{\xi}_{mnk}\right)\nonumber\\
& \leq {v_{mnk}} ,\forall\;\;\boldsymbol{\xi}_{mnk}^{H} \mathbf{R}_{mnk} \boldsymbol{\xi}_{mnk} \leq 1, \forall m \neq n, k, \label{subeq36c}\\
& \mathbf{G}_{nk} \succeq {\bf 0}.
\end{alignat}
\end{subequations}
Even though the constraints in \eqref{eq36} above are convex, it is mathematically intractable to evaluate the optimal TPC due to the presence of infinitely many SINR constraints, namely one for each value of $\boldsymbol{\xi}_{mnk}$. However, these infinitely many constraints can be reduced into a few constraints by employing the S-lemma \cite{boyd2004convex}, as given below.
\begin{lemma}
Let $\mathbf{X}$, $\mathbf{Y}$ $\in \mathbb{C}^{N \times N}$ denote complex Hermitian matrices, and $\mathbf{a} \in \mathbb{C}^N $, $\mathbf{b} \in \mathbb{C}^N$ and $d \in \mathbb{R}$. The following condition
\begin{align*}
{\bf a}^{H} {\bf Xa} +{\bf b}^{H}{\bf a}+{\bf a}^{H}{\bf b} + d \geq 0, \ \ \forall\ {\bf a}^{H} {\bf Ya} \leq 1,
\end{align*}
holds true if and only if there exists a value of $\lambda \geq 0$ so that $\left[\begin{matrix}
 \mathbf{X} + \lambda \mathbf{Y} & \mathbf{b}\\\mathbf{b}^H & d - \lambda
\end{matrix}\right]\succeq \mathbf{0}.$
%\begin{equation*}
%\left[\begin{matrix}
% \mathbf{X} + \lambda \mathbf{Y} & \mathbf{b}\\\mathbf{b}^H & d - \lambda
%\end{matrix}\right]\succeq \mathbf{0}.
%\end{equation*}
\end{lemma}
In order to apply the S-lemma, set $\mathbf{x} = \boldsymbol{\xi}_{mnk}$ and $\mathbf{B} = \mathbf{R}_{mnk}$. The constraints in \eqref{subeq36b} and \eqref{subeq36c} can be recast as the linear matrix inequalities, which are given as
\begin{align}
    & \boldsymbol{\Phi}_{nk}\left({\{\mathbf{G}_{ni}\}}_{i=1}^{K},{\{v_{mnk}\}}_{m},\lambda_{nnk}\right)  \triangleq \boldsymbol{\Phi}_{nk} \succeq{\bf 0}, \nonumber \\& \boldsymbol{\Psi}_{mnk}\left({\{{\bf G}_{mi}\}}_{i=1}^{U},v_{mnk},\lambda_{mnk}\right)\triangleq \boldsymbol{\Psi}_{mnk} \succeq{\bf 0},\label{subeq39acc}
\end{align}
where the matrices above are defined as
%New equation
\begin{align}\label{eq37}
&\boldsymbol{\Phi}_{nk}\triangleq \left[ \begin{matrix}
\mathbf{I}\\ \hat{\bf h}_{nnk}^H \end{matrix}\right] \left({1\over \gamma_{nk}}{\mathbf{G}}_{nk}-\sum\limits_{i\neq k}^{K}{\mathbf{G}}_{ni}\right) \left[\begin{matrix}
\mathbf{I}&\hat{\bf h}_{nnk}\end{matrix}\right]\nonumber\\
&+ \left[\begin{matrix}
\lambda_{nnk}\mathbf{R}_{nnk} & {\bf 0} \\
{\bf 0} & -\sigma_{nk}^{2}-\sum\limits_{m\neq n}^{N}{v}_{mnk}-\lambda_{nnk}\end{matrix}\right],
\end{align}
%\begin{comment}
    \begin{align}\label{eq38}
&\!\!\!\!\!\!\!\!\!\!\!\!\!\!\! \boldsymbol{\Psi}_{mnk}\triangleq \left[ \begin{matrix}
{\mathbf{I}}\\\hat{\bf h}_{mnk}^H \end{matrix}\right] \left(-\sum\limits_{m\neq n}^{N}{\mathbf{G}}_{mk}\right)\left[\begin{matrix}
{\mathbf{I}}&\hat {\bf h}_{mnk} \end{matrix}\right]\nonumber\\
&+ \left[\begin{matrix}
\lambda_{mnk}{\mathbf{R}_{mnk}} & {\bf 0}\\ {\bf 0} & {v}_{mnk}-\lambda_{mnk}\end{matrix}\right].
\end{align}
%\end{comment}
\begin{comment}
    \begin{figure*}
   \begin{align}\label{eqqw37}
&\boldsymbol{\Phi}_{nk}\left({\{{\bf G}_{ni}\}}_{i=1}^{K},{\{v_{mnk}\}}_{m},\lambda_{nnk}\right)
\triangleq \left[ \begin{matrix}
\mathbf{I}\\ \hat{\bf h}_{nnk}^H \end{matrix}\right] \left({1\over \gamma_{nk}}{\mathbf{G}}_{nk}-\sum\limits_{i\neq k}^{K}{\mathbf{G}}_{ni}\right) \left[\begin{matrix}
\mathbf{I}&\hat{\bf h}_{nnk}\end{matrix}\right]+ \left[\begin{matrix}
\lambda_{nnk}\mathbf{R}_{nnk} & {\bf 0} \\
{\bf 0} & -\sigma_{nk}^{2}-\sum\limits_{m\neq n}^{N}{v}_{mnk}-\lambda_{nnk}\end{matrix}\right],
\end{align}
\begin{align}\label{eq3y8}
& \boldsymbol{\Psi}_{mnk}\left({\{{\bf G}_{mi}\}}_{i=1}^{K},v_{mnk},\lambda_{mnk}\right) \triangleq \left[ \begin{matrix}
{\mathbf{I}}\\\hat{\bf h}_{mnk}^H \end{matrix}\right] \left(-\sum\limits_{m\neq n}^{N}{\mathbf{G}}_{mk}\right)\left[\begin{matrix}
{\mathbf{I}}&\hat {\bf h}_{mnk} \end{matrix}\right] + \left[\begin{matrix}
\lambda_{mnk}{\mathbf{R}_{mnk}} & {\bf 0}\\ {\bf 0} & {v}_{mnk}-\lambda_{mnk}\end{matrix}\right].
\end{align}
\end{figure*}
\end{comment}
The optimization problem for the robust centralized beamformer design can now be recast as:
 \begin{subequations}\label{eq39}
\begin{alignat}{5}
&\min_{\{{\bf G}_{nk}\},\{\lambda_{mnk}\},\{v_{mnk}\}}\sum_{n=1}^{N}\beta_{n}\left(\sum_{k=1}^{K}{\text{ Tr}}({\bf G}_{nk})\right) \nonumber\\
& \text{s.t.}\;\;\boldsymbol{\Phi}_{nk}\succeq{\bf 0},\ \boldsymbol{\Psi}_{mnk}\succeq{\bf 0}, \forall m\neq n,\label{subeq39c}\\
&\;\;\;\;\;\;\;{\bf G}_{nk}\succeq{\bf 0},\ \ \textrm{and} \ \  \lambda_{mnk}\geq 0, \forall n, k, m.\label{subeq39d}
\end{alignat}
\end{subequations}
Note that the optimization problem in \eqref{eq39} is a SDP that can be efficiently evaluated similar to \eqref{eq9} described in Section \ref{MainSysModel}. Following this, the principal eigenvector of the matrix $\mathbf{G}_{nk}$ having a unit-norm can be chosen as the optimal solution $\mathbf{g}_{nk,\mathrm{opt}}$, and the associated hybrid TPC can be designed by employing the BL-method developed in Section \ref{MainSysModel}. The ADMM-based robust asynchronous distributed beamformer (R-ADBF) design relying on  imperfect CSI knowledge is described next.
\subsection{Robust ADBF design}\label{RDMCMU}
%Define the following two  auxiliary variables, akin to the perfect CSI model described in Section \ref{SDMCMU} as
 %\begin{equation}\label{eq40}
%{V_{nk}}=\sum_{{m\neq n}}^{N}{ v_{mnk}} \ \ \forall n, k,
%\end{equation}
%where $v_{mnk}$ denotes the worst-case ICI power inflicted by $\mathrm{BS}_{m}$ upon $\mathrm{UE}_{nk}$ and ${V_{nk}}$ denotes the total ICI power received at $\mathrm{UE}_{nk}$ from the all the BSs. 
To begin with, one can interchange the subscripts $m$ and $n$ in $\boldsymbol{\Psi}_{mnk}$ without changing the original problem, therefore, the matrix $\boldsymbol{\Psi}_{mnk}$ can be rewritten as $\boldsymbol{\Psi}_{nmk}\left({\{{\bf G}_{ni}\}}_{i=1}^{K},v_{nmk},\lambda_{nmk}\right)\succeq{\bf 0},\forall  m\neq n,k$.
\begin{comment}
     \begin{subequations}\label{eq41}
\begin{alignat}{6}
&\min_{\{{\bf G}_{nk}\succeq{\bf 0}\},\atop\{\lambda_{mnk}\geq 0\},\{v_{mnk}\}}\ \ \ \ \ \ \  \sum_{n=1}^{N}\sum_{k=1}^{{K}}\beta_{n}{\text{Tr}}({\bf G}_{{nk}}) \label{subeq41a}\\
&\ \ \ \ \ \ \ \text{s.t.}\ \ \boldsymbol{\Phi}_{nk}\left({\{{\bf G}_{ni}\}}_{i=1}^{K},V_{nk},\lambda_{nnk}\right)\succeq{\bf 0},\forall n,k,\label{subeq41b}\\
&\ \ \ \ \ \ \ \ \ \ \ \ \boldsymbol{\Psi}_{nmk}\left({\{{\bf G}_{ni}\}}_{i=1}^{K},v_{nmk},\lambda_{nmk}\right)\succeq{\bf 0},\forall  m\neq n,k, \label{subeq41c}\\
%&\ \ \ \ \ \ \ \ \ \ \ \ \sum_{k=1}^{K}{\text{Tr}}({\bf F}_{nk})=p_{n},\forall n,\label{subeq41d}\\
&\ \ \ \ \ \ \ \ \ \ \ \ \ {\bf G}_{nk}\succeq{\bf 0},\forall n,k,\label{subeq41e}\\
&\ \ \ \ \ \ \ \ \ \ \ \ \ \lambda_{mnk}\geq 0,\forall  m\neq n,k.\label{subeq41f}
\end{alignat}
\end{subequations}
\end{comment}
Following this, the constraints in \eqref{eq39} can be decomposed into $N$ independent convex sets as
\begin{align}\label{eq42}
&{\cal C}_{n}\!=\!\Bigg \lbrace \left({\{{\bf G}_{nk}\}}_{k},{\{\lambda_{nmk}\}}_{m,k},{\{  V_{nk}\}}_{k},{\{v_{nmk}\}}_{m,k}\right)\vert
\nonumber\\
& \hspace{1cm} \boldsymbol{\Phi}_{nk}\succeq\!{\bf 0},\forall k, \ \ \boldsymbol{\Psi}_{nmk} \succeq\!{\bf 0}, \forall m \neq n, k,\nonumber\\
& \hspace{0.2cm} \lambda_{nmk}\geq 0,\  {\bf G}_{nk}\!\succeq\!{\bf 0},  \ V_{nk}\geq 0, \ \forall m, n, k, \Bigg \rbrace,\forall n.
\end{align}
The optimization problem in \eqref{eq39} can now be reformulated as
\begin{align}\label{eq43}
   &\min_{\{{\bf G}_{nk}\},\{\lambda_{nmk}\},\atop{\{{\bf v}_{n}\},{\bf v}}}\sum_{n=1}^{N}\beta_{n}\sum_{k=1}^{{K}}{\text{Tr}}({\bf G}_{{nk}})\nonumber \\
& \hspace{-0.3cm} \text{s.t.} \left({\{{\bf G}_{nk}\}}_{k},{\{\lambda_{nmk}\}}_{m,k},{\bf v}_{n}\right)\in{\cal C}_{n}, {\bf v}_{n}={\bf W}_{n}{\bf v}, \forall n, 
\end{align}
which is similar to problem \eqref{eq18}. Therefore, Algorithm \ref{SDAD} can now be readily applied for the design of the SDBF in this scenario having CSI uncertainty, Furthermore, the ADBF design can be carried out by applying Algorithm \ref{ADCU} and Algorithm \ref{ADBS}. Following this, the FD ADBF can be decomposed into the RF and BB TPC for obtaining the associated hybrid ADBF via the BL-method developed in Section \ref{MainSysModel}.  
Using Algorithm \ref{SDAD}, each BS iteratively approaches the optimal solution until the pertinent ICI information $\left({\bf v}_{n}^{(i+1)}\right)$ for $\mathrm{BS}_n$ is cancelled from the ICI information $\left({\bf v}^{(i+1)}\right)$, formulated ${\bf W}_{n}{\bf v}^{(i+1)}={\bf v}_{n}^{(i+1)}$, for all $n$. It is important to note that the quantities $\{\mathbf{G}_{nk}\}$ and $\{\lambda_{nmk}\}$ obtained in Step-4 of Algorithm 2 may not be feasible for the primal problem \eqref{eq43}. This is due to the fact that the ADMM algorithm works in the dual domain, which does not guarantee the constraint $\mathbf{W}_n \mathbf{v}^{(i+1)} = \mathbf{v}_n^{(i+1)}$ to hold true prior to reaching convergence. Nevertheless, each BS can perform additional optimization, as shown in the problem below

\begin{alignat}{5}\label{eq44}
&\min_{\{{\bf G}_{nk} \succeq \mathbf{0} \}_{k},\{\lambda_{mnk} \geq 0 \}_{n,k}}\sum_{n=1}^{N}\beta_{n}\left(\sum_{k=1}^{K}{\text{ Tr}}({\bf G}_{nk})\right)\nonumber \\
& \hspace{1cm} \text{s.t.} \ \boldsymbol{\Phi}_{nk}\succeq{\bf 0},  \ \ \boldsymbol{\Psi}_{nmk} \succeq{\bf 0}, \forall m\neq n, %\nonumber \\
%&\hspace{1.5cm} \boldsymbol{\Psi}_{nmk}\left({\{{\bf G}_{ni}\}}_{i=1}^{K},v_{nmk},\lambda_{nmk}\right)\succeq{\bf 0}, \forall m\neq n,
\end{alignat}
by employing the tentatively consented ICI power vector $\mathbf{v}^{(i+1)}$. The quantities $\{\mathbf{G}_{nk}\}$ and $\{\lambda_{nmk}\}$ are feasible for the SDR problem \eqref{eq44}, provided that the optimization problem \eqref{eq44} yields feasible solutions for all the BSs. If at least one BS declares the infeasibility of \eqref{eq44}, additional iterations of Algorithm \ref{ADCU} are required for convergence, as it may not have reached a reasonable consensus regarding the global ICI vector $\mathbf{v}^{(i+1)}$.
The overhead of backhaul signaling required for the centralized and ADBF design schemes can be determined as follows. In the centralized TPC design, the global CSI for each BS is obtained by exchanging the local CSI of each BS through backhaul links. The total signaling overhead for scalar-valued complex channel coefficients in this case is proportional to $2N_{t}KN\left(N-1\right)$. 
%For the ADBF design, in each iteration of Algorithm \ref{ADCU}, the local  variable updates $(\mathbf{v}_n-\mathbf{\nu_n}) \in \mathbb{C}^{NK \times 1}$ are transmitted by each BS to the CU, which incurs a total signaling overhead of $NK$.  It can be seen that the distributed TPC design results in significant reduction of the backhaul signalling.
Furthermore, in \cite{jafri2022robust}, the overall signalling overhead for the $n$th BS at any iteration is given as $(N-1)NK$. However, for the proposed ADBF design algorithm, in each iteration of Algorithm \ref{ADCU}, the local variable updates $\left({\bf v}_{{n}}(i+1)- \frac{1}{c}\boldsymbol{\nu}_{n}(i)\right) \in \mathbb{C}^{NK \times 1}$ are transmitted by each BS to the CU, which incurs a total signaling overhead of $NK$. Therefore, the signalling overhead is significantly reduced in comparison to \cite{jafri2022robust} and the centralized TPC.

%Due to lack of space, the detailed derivations of the computational complexity had to be relegated to our technical report in \cite{TechReport}. As described in \cite{TechReport}, 
A brief analysis of the computational complexity of the proposed ADBF design and BL scheme is presented next. The complexity of each scheme is quantified in terms of complex additions and multiplications. Table-\ref{tab:SBL} and Table-\ref{tab:MSIP} details the computational cost of the various steps of the BL algorithm and ADBF scheme, respectively.
One can observe that the FD ADBF design incurs a complexity of order $\mathcal{O}\left(N_{t}^{3}\right)$. Next, the FD TPC is decomposed into its constituent RF and BB TPCs using the BL algorithm. This leads to a computational complexity of order $\mathcal{O}\left(G^3\right)$ due to the matrix inversion of size $G \times G $ in \eqref{eq. mu and sigma}. Since $G >> N_t$, the overall complexity of the ADBF design can be closely approximated by $\mathcal{O}\left(G^3\right)$. On the other hand, the complexity of state-of-the-art hybrid TPC design method, i.e., the simultaneous orthogonal matching pursuit (SOMP) algorithm \cite{alkhateeb2015limited} is of the order of $\mathcal{O}\left(N_t KG\right)$. However, the SOMP algorithm has a poor performance in comparison to the BL-based approach, since the performance of the SOMP algorithm is highly sensitive both to the choice of the dictionary matrix and to the stopping criterion.
% BL scheme table
\begin{table*}
    \centering
    \caption{Complexity of the BL scheme per-EM-iteration}\label{tab:SBL}
\begin{tabular}{|c|c|c|}
    \hline
 Operation & Complex Multiplications  & Complex Additions  \\ [0.5ex]
 \hline
 
$\mathbf{\Omega}^{(j)}$& \makecell{$\dfrac{G^3}{2} + \dfrac{3G^2}{2} + \dfrac{N_{t}G(G+1)}{2} + G^2$} & \makecell{$\dfrac{G^3}{2} - \dfrac{3G^2}{2} + \dfrac{(N_{t}-1)G(G+1)}{2} + G^2$} \\
\hline

$\tilde{\mathbf{G}}_{\mathrm{BB},n}^{(j)}$& $GN_{t}K+ GK$ & $GK(N_{t}-1)$ \\
 \hline
$\widehat{\gamma}_i^{(j)}$ &$KN_{\mathrm{RF},n}$ &$G^2 + K$\\
 \hline
\end{tabular}
\end{table*}
%................................
%.... ADBF scheme table
\begin{table*}
    \centering
    \caption{Complexity of ADBF design per-iteration}\label{tab:MSIP}
\begin{tabular}{|c|c|c|}
    \hline
 Operation & Complex Multiplications  & Complex Additions  \\ [0.5ex]
 \hline
\makecell{Algorithm 3:\\ Step-3} & \makecell{$3N_t^3 + \left(NK\right)^2 \left(N - 1\right) + 2NK$} & \makecell{$ NK \left(N \left(N - 1\right)K - 1 \right) + 3NK$\\
$ + 3N_t^2 \left(N_t-1\right) + 2N_tK $} \\ [0.5ex]
\hline
\makecell{Algorithm 2:\\ Step-8} & \makecell{$ \frac{3N^4K^3\left(N_c - 1\right)^2}{2} + \frac{N^3K^3\left(N - 1\right)^3}{2} + \frac{3N^2K^2\left(N - 1\right)^2}{2} $\\ $ +  2N^3K^2\left(N - 1\right) $ } & \makecell{$ \frac{N^4K^3\left(N - 1\right)^2}{2} + \frac{N^3K^3\left(N - 1\right)^3}{2} - \frac{N^2K^2\left(N - 1\right)^2}{2} $\\ $ +  N^4K^2\left(N - 1\right)^2 + N^4 K^2$} \\ [0.5ex]
 \hline
\makecell{Algorithm 3:\\ Step-5} &$\left(NK\right)^2 \left(N - 1\right)$ & $2NK$\\
 \hline
\end{tabular}
\end{table*}
\section{Convergence Analysis of the ADBF Design Algorithm}\label{converg}
This section discusses the convergence behavior of the proposed ADBF design. In general, the arrival of updates at the CU is random in nature and depends on the number of BSs participating in the coordinated beamformer design. Hence, we assume that at any CU in iteration $i$, the updates from all the $N$ BSs have an equal probability of arriving at the CU. Let us assume that the CU clock $i$ and each BS clock $i_n$ runs for $Q$ and $Q_n $ iterations, respectively. The individual optimization problem constructed for determining the beamformer at each $\mathrm{BS}_n$ is given as follows
\begin{align} 
 f_n({\bf v}_{n}^{(i_n+1)},{\bf G}_{nk})&= \arg\min \Bigg\{\sum_{k=1}^{{K}}\beta_{n}{\text{Tr}}({\bf G}_{{nk}}) \nonumber \\ &  +{c^{(i_n)}\over 2}\left\Vert{\bf W}_{n}{\tilde{\bf v}(n)}-{\bf v}_{n}\right\Vert^{2}-{\boldsymbol{\nu}_{n}^{(i_n)}}^T{\bf v}_{n}\Bigg\},\nonumber \\
& \text{s.t.} \ \ \ \left({\big\{{\bf G}_{nk}\big\}}_{k},{\bf v}_{n}\right)\in{\cal C}_{n},
\end{align}
Since, the information exchange between each $\mathrm{BS}_n$ and the CU is in terms of the local ICI ${\bf v}_{n}$ and global ICI ${\bf v}$, one can simplify $f_n({\bf v}_{n}^{(i_n+1)},{\bf G}_{nk})$ as $f_n({\bf v}_{n}^{(i_n+1)})$ . 
 Let $\Bar{\bf v}_{n}= \frac{1}{Q_n}\sum_{i_n=1}^{Q_n}{\bf v}_{n}^{i_n}$ denote the average of all the local ICI updates ${\bf v}_{n}$ generated throughout $Q_n$ iterations by all the BSs.
Let $\Bar{\bf v}$ denote the average of all the global ICI updates ${\bf v}^{(i)}$ generated by CU throughout its $Q$ iterations. Next, we demonstrate that the ADBF design algorithm converges with the order of $\mathcal{O}(\frac{N \tau}{QS})$.
%In \cite{zhang2014asynchronous}, the authors provided convergence rate for a more generalized asynchronous distributed implementation of a convex problem in Section IV.  
%The equivalent convergence rate for solving our optimization problem using ADBF design is given as follows:
\begin{lemma}
Let $({\bf v}_{n}^*,{\bf v}^*)$ be the optimal primal solution, and $\left\{{\boldsymbol{\nu}_{n}^*}\right\}_{n=1}^{N}$ the corresponding optimal dual solution. It follows that
\begin{align}
&{\mathbb{E}}\bigg[ \sum_{n=1}^{N} f_n(\Bar{\bf v}_{n})-f_n({{\bf v}_{n}^*})+ \langle \boldsymbol{\nu}_{n}^*,\Bar{\bf v}_{n} -{\bf W}_{n}\Bar{\bf v} \rangle \bigg] \nonumber \\ & \leq 
    \frac{N\tau}{2QS}\Bigg\{ \sum_{n=1}^{N}c\norm{{\bf v}(n)^0-{\bf v}*}^2+\frac{1}{c}\norm{\boldsymbol{\nu}_{n}^0 - \boldsymbol{\nu}_{n}^*}^2 \Bigg\}
\end{align}
\end{lemma}
where ${\bf v}(n)^0$ and $\boldsymbol{\nu}_{n}^0$ are the initial values of the variables ${\bf v}(n)$ and $\boldsymbol{\nu}_{n}$, respectively, at the $\mathrm{BS}_n$.\\
\textit{Proof:} To obtain the upper bound of the convergence rate of ADBF design algorithm, we consider the following worst-case conditions:
\begin{itemize}
    \item Only $S$ BS updates out of $N$ are received at CU in any iteration.  \item The probability that the update from BS $n$ belongs to the set of $S$ updates received at CU in iteration $i+1$ $\mathcal{A}_{i+1}$ is $\frac{S}{N}$.
    \item The CU receives each BS updates only once every $\tau$ iterations. As a result, each BS runs for only $\frac{Q}{\tau}$ iterations. Therefore, $\bar{\bf {v}}_{n}=\frac{\tau}{Q} \sum_{i_{n}=0}^{\frac{Q}{\tau}-1} {\bf {v}}_{n}^{i_{n}+1}$.
\end{itemize}
Note that $\bar{\bf {v}}=\frac{1}{Q} \sum_{i=0}^{Q-1} {\bf {v}}^{i+1}$ and ${\bf {v}}^{i+1} 
={\bf W}^{\dagger}\left(\tilde{\bf v}^{(i+1)}-{1\over {c}}\tilde{\boldsymbol{\nu}}^{(i)}\right)$, where   $\tilde{\bf v}^{(i+1)}=\left[\left(\bf\widehat{v}_{1}\right)^{T},\ldots,\left({\bf \widehat{v}}_{N}\right)^{T}\right]^{T}$ and 
$\tilde{\boldsymbol{\nu}}^{(i)}=\left[\left(\boldsymbol{\widehat{\nu}}_{1}^{(i)}\right)^{T},\ldots,\left(\boldsymbol{\widehat{\nu}}_{N}^{(i)}\right)^{T}\right]^{T}$,
where $\mathbf{\widehat{v}}_n,\mathbf{\widehat{\nu}}_n$ are the most recent updates received from the $n$th BS at the CU. Therefore, $\Bar{\bf {v}}= \frac{1}{Q} \sum_{i=0}^{Q-1}{\bf W}^{\dagger}\left(\tilde{\bf v}^{(i+1)}-{1\over {c}}\tilde{\boldsymbol{\nu}}^{(i)}\right)$. However, observe that each BS updates involved in $\bar{\bf {v}}$ is repeated $\tau$ times. Therefore, $\Bar{\bf {v}}$ can be written as an average of global ICI updates over $\frac{Q}{\tau}$ iterations, where each CU iteration will receive distinct updates from all the BSs, i.e.,
\begin{align}
    \bar{\bf {v}}=\frac{\tau}{Q} \sum_{i=0}^{\frac{Q}{\tau}-1} {\bf {v}}^{i+1}.
\end{align}
The convergence equation for the ADBF design algorithm can be written as \cite{converg, async}
\begin{align}
   & \mathbb{E}\Biggl[\sum_{n=1}^{N} \sum_{i=0}^{Q-1}{\mathrm{Pr}\left(n \in A_{i+1}\right)}\biggl\{f_{n}\left({\bf {v}}_{n}\right)-f_{n}\left(\bf {v}^{*}\right) \nonumber \\ & +\left(\boldsymbol{\nu}_{n}^{*}\right)^T({\bf {v}}_{n}-{\bf {v}})\biggr\} \Biggr] \leq \frac{\theta}{2},
   \label{ConverEq}
\end{align}
where $\theta = \left\{\sum_{n=1}^{N} c\left\|{\bf {v}}(n)^{0}-\bf {v}^{*}\right\|^{2}+\frac{1}{c}\left\|\boldsymbol{\nu}_{n}^{0}-\boldsymbol{\nu}_{n}^{*}\right\|^{2}\right\}$ and the quantity $\mathrm{Pr}\left(n \in A_{i+1}\right)$ denotes the probability that the update from BS $n$ belongs to the set $A_{i+1}$. Since, each BS runs only for $\frac{Q}{\tau}$ iterations, one can rewritten the expression \eqref{ConverEq} as 
\begin{align}
   & \mathbb{E}  \Biggl[\sum_{n=1}^{N} \sum_{i=0}^{\frac{Q}{\tau}-1}{\frac{S}{N}}\biggl\{ f_{n}\left({\bf {v}}_{n}^{i+1}\right) - f_{n}\left(\bf {v}^{*}\right) \nonumber \\ & + \left(\boldsymbol{\nu}_{n}^{*}\right)^T\left({\bf {v}}_{n}^{i+1}-{\hat{\bf {v}}}^{i+1}\right)\biggr\}\Biggr] \leq \frac{\theta}{2}.
\end{align}
Since $f_n\left({\bf {v}}_n\right)$ are convex functions for each $n$, following the Jensen's inequality one can write $\sum_{i=0}^{\frac{Q}{\tau}-1}f_n\left({\bf {v}}_{n}^{i+1}\right) \geq \frac{Q}{\tau}f_n(\bar{\bf {v}}_n) $. Hence, the above equation reduces to
\begin{equation}
   {\frac{S}{N}} \frac{Q}{\tau}\mathbb{E}  {\left[\sum_{n=1}^{N} \left \{f_{n}\left(\bar{\bf {v}}_{n}\right)-f_{n}\left(\bf {v}^{*}\right)+(\boldsymbol{\nu}_{n}^{*})^T(\bar{\bf {v}}_{n}-\bar{\bf {v}})\right\}\right] } \leq \frac{\theta}{2}
\end{equation}
\begin{equation}
   \mathbb{E}  {\left[\sum_{n=1}^{N} \left \{f_{n}\left(\bar{\bf {v}}_{n}\right)-f_{n}\left(\bf {v}^{*}\right)+(\boldsymbol{\nu}_{n}^{*})^T(\bar{\bf {v}}_{n}-\bar{\bf {v}})\right\}\right] } \leq {\frac{N\tau}{SQ}}\frac{\theta}{2}.
\end{equation}
Therefore, the convergence rate of the proposed asynchronous algorithm is given as $\mathcal{O}\left( \frac{N \tau}{QS}\right)$. which  can be intuitively explained as follows:
\begin{itemize}
\item When the number of coordinated cells $N$ is large, the specific fraction of information shared by each BS is reduced. This corresponds to the situation wherein the update from each BS is less influential. Hence, the number of iterations required for convergence increases.
\item A large $S$ corresponds to the scenario that information from a large fraction of BSs is incorporated in each CU update in the design of the distributed beamformer. Therefore, the number of iterations required for convergence of the ADBF algorithm decreases upon increasing $S$.
\item Recall that updates from each BS are incorporated by the CU at least $\frac{Q}{\tau}$ times in $Q$ CU iterations. Hence, a large $\tau$ implies that information from the slower BSs is not utilized by the CU frequently. Thus, the iterations required for convergence increases upon increasing $\tau$.
\end{itemize}
%In , the authors have given detailed proofs for the convergence conditions of similar asynchronous distributed ADMM method and also provided simulation results to support the convergence rate. The order of $\mathcal{O}(\frac{N \tau}{QS})$ convergence rate can be intuitively explained as follows. We have provided simulation results that verifies the given convergence rate of ADBF design. Also, when $S = N$ or $\tau = 1$ the proposed ADBF design reduces to SDBF design and the resultant convergence rate is $\mathcal{O}(\frac{1}{Q})$ which is same as the convergence rate of ADMM. \cite{admm_conv}
\begin{figure*}
\centering
\subfloat[]{\includegraphics[width=0.5\linewidth]{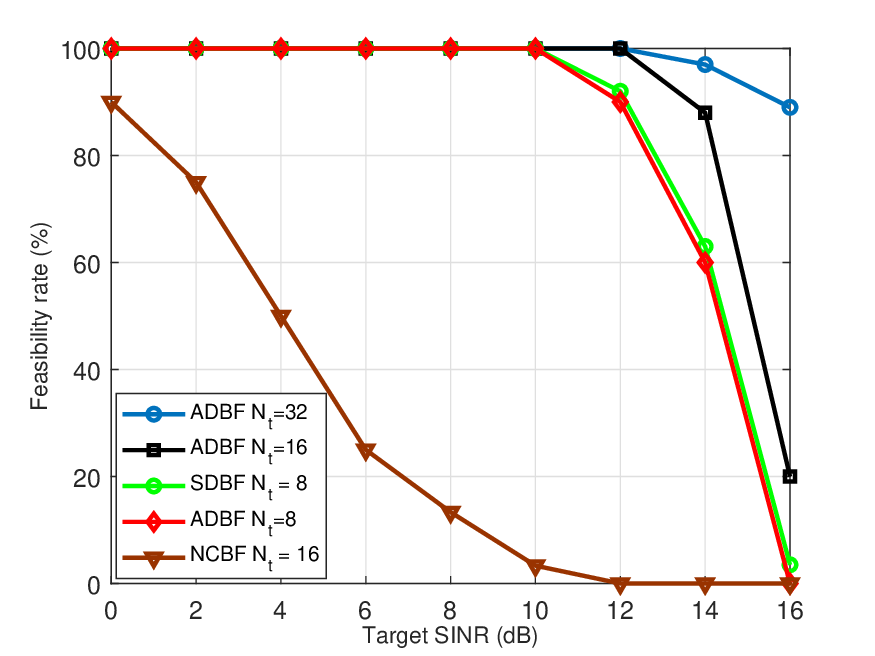}\label{f2a}}
\subfloat[]{\includegraphics[width=0.5\linewidth]{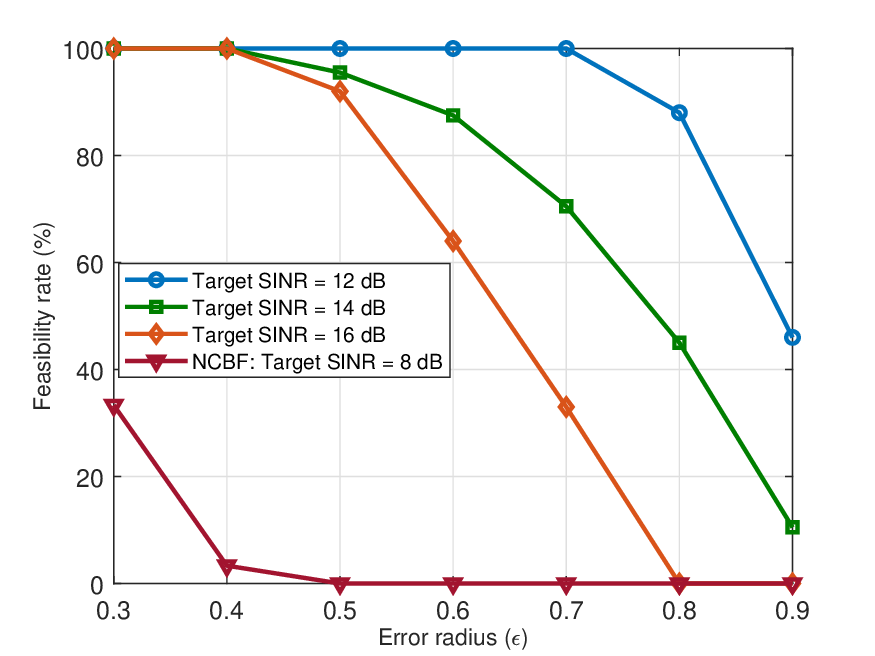}\label{f1b}}
%\subfloat[]{\includegraphics[width=0.34\linewidth]{Figure5.eps}\label{f4bba}}
\caption{(a) Feasibility rate versus target SINR parameterized by different number of BS TAs for $\epsilon = 0.4$; (b) Feasibility rate versus error radius for $N_t = 16$.}\label{f8}
\end{figure*}
%............Next figure
\begin{figure*}
\centering
\subfloat[]{\includegraphics[width=0.5\linewidth]{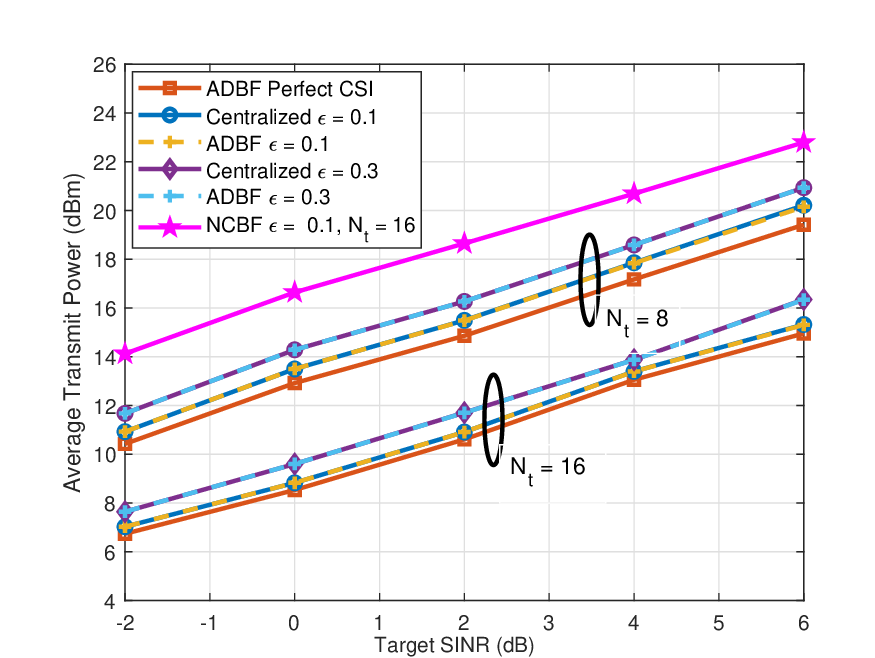}\label{f111b}}
\subfloat[]{\includegraphics[width=0.5\linewidth]{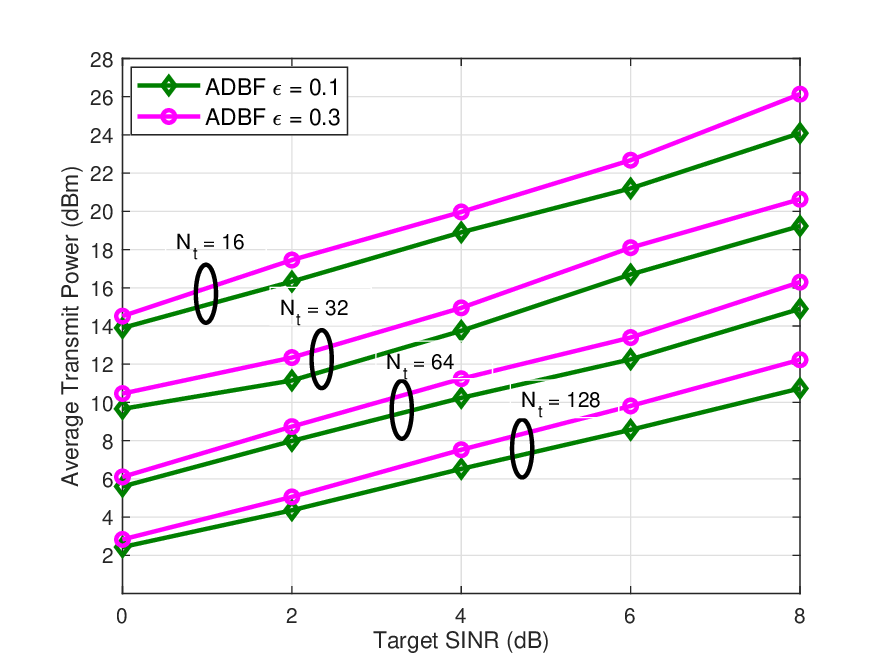}\label{f4bbaa}}
\caption{Average transmit power versus target SINR for; (a) $N=2$, $K = 2$, $S = 1$; (b) $N=4$, $K = 3$, $S = 3$.}\label{f88}
\end{figure*}
% Transmit power vs Channel realization plots
\begin{figure*}
\centering
\subfloat[]{\includegraphics[width=0.5\linewidth]{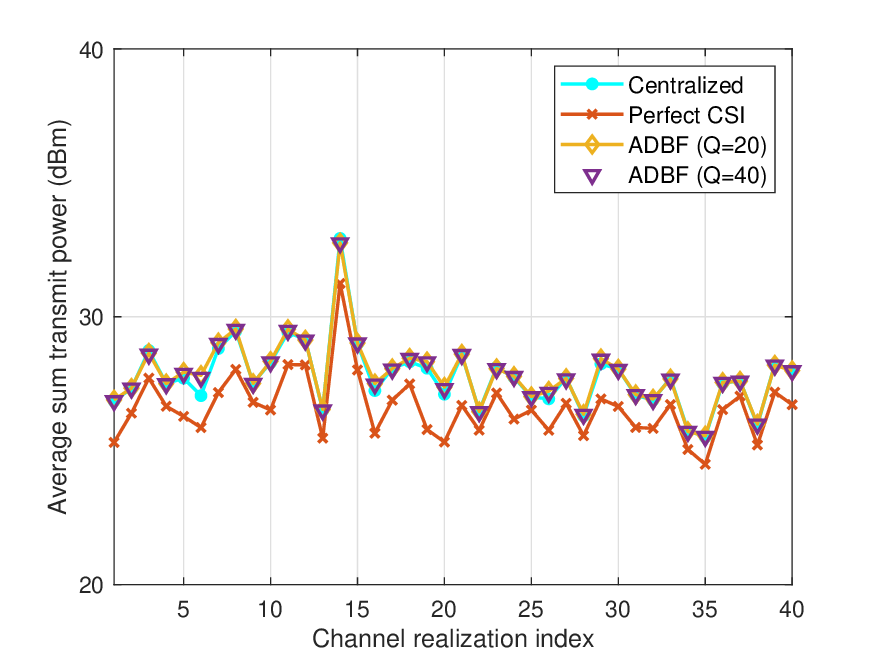}\label{f6a}}
\subfloat[]{\includegraphics[width=0.5\linewidth]{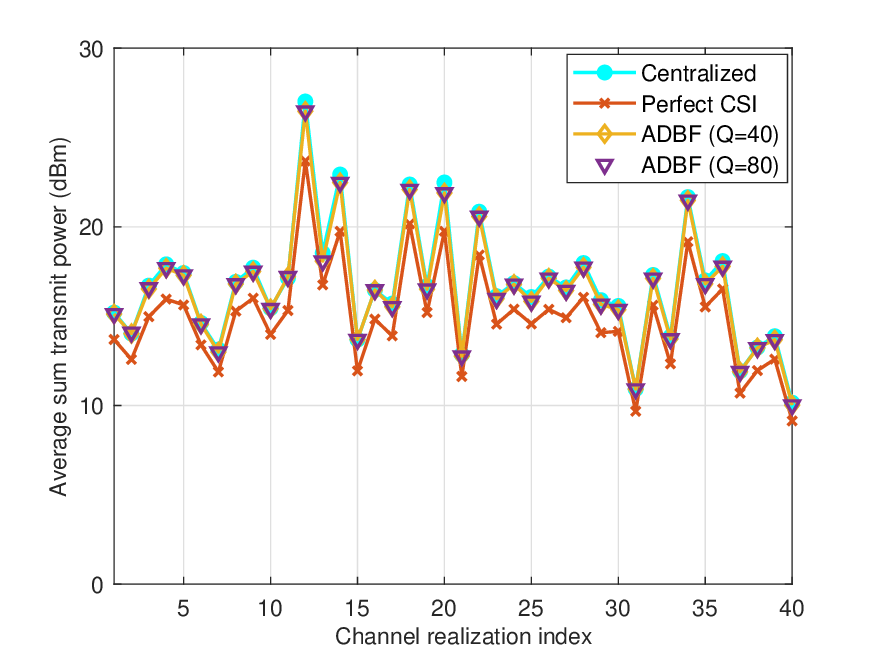}\label{f6b}}
\caption{Sum power comparison between centralized and ADBF designs for $S= 1$, $K = 2$, $\tau = 4$, $N_t = 16$, $\epsilon = 0.1$:} (a) $N=2$; (b) $N=3$.\label{f6}
\end{figure*}
%Power Accuracy plots
\begin{figure*}
\centering
\subfloat[]{\includegraphics[width=0.33\linewidth]{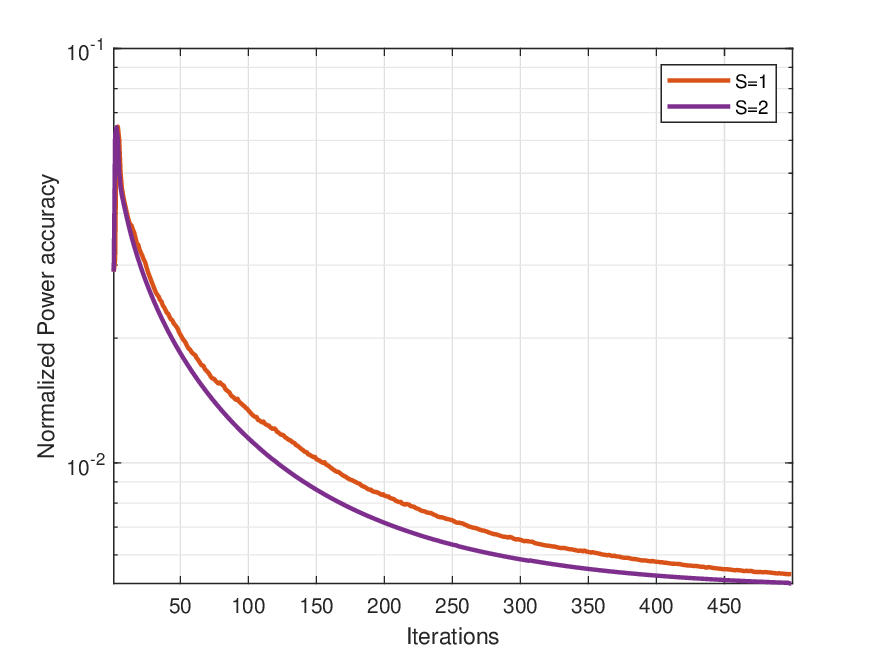}\label{f5a}}
\subfloat[]{\includegraphics[width=0.33\linewidth]{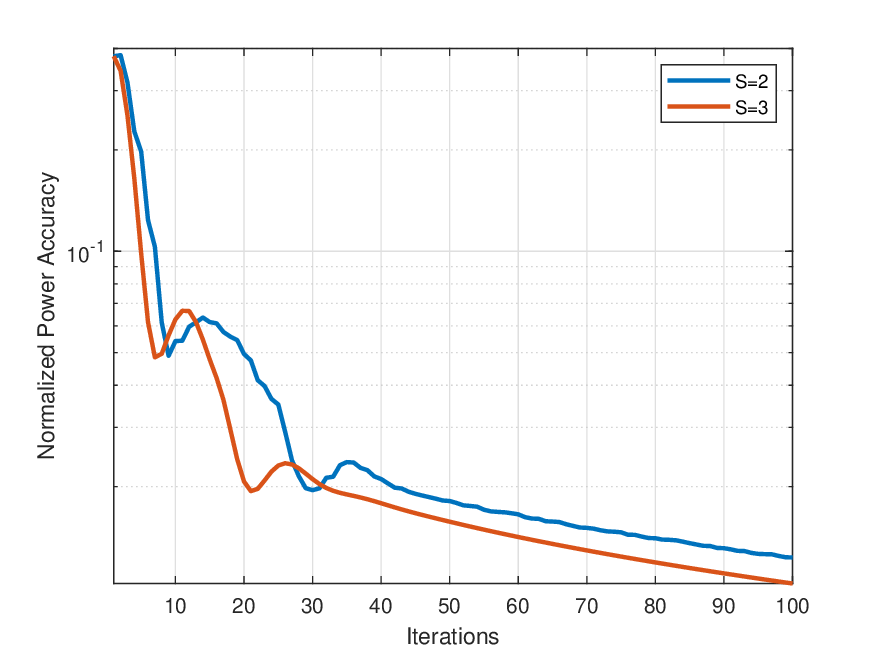}\label{f5b}}
\subfloat[]{\includegraphics[width=0.33\linewidth]{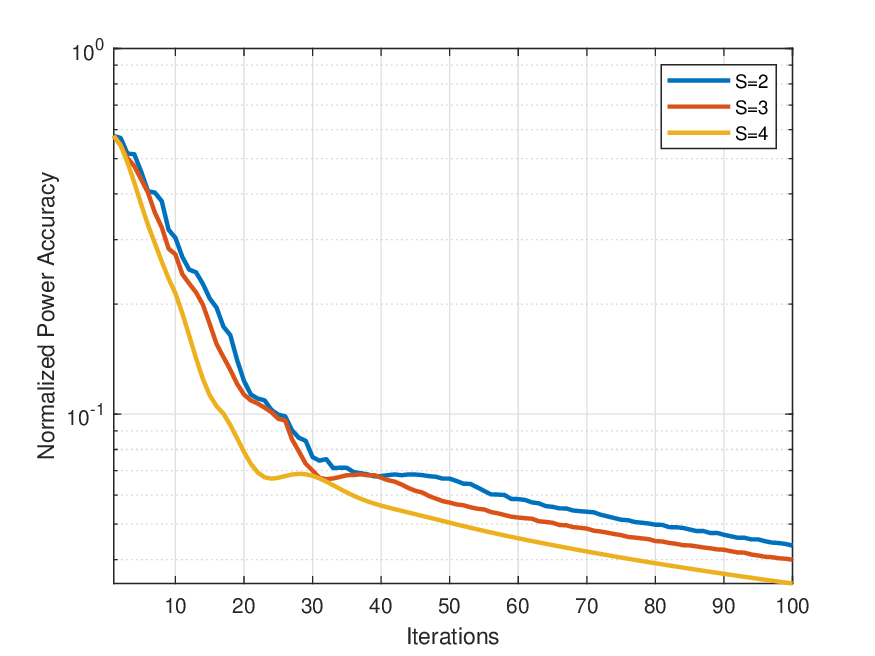}\label{f5c}}
\caption{Normalized power accuracy versus the number of iterations for $\tau = 4$, $\epsilon = 0.1$ and $N_t = 16$: (a) $N=2$, $K=2$; (b) $N=3$, $K=2$; (c) $N=4$, $K=2$.}\label{f5}
\end{figure*}
%............Simulation results
\section{Simulation Results}\label{sim}
This section characterizes the performance of the proposed asynchronous distributed hybrid TPC design considering both perfect as well as imperfect CSI for mmWave MCC networks. In the simulation model, we consider an $N$-cell network with each BS equipped with $N_t$ TAs, and $K$ users equipped with a single RA each. The number of RFCs $N_{\mathrm{RF},n}$ at each BS in a cell is set equal to the number of users served by that BS. The gains of the multipath components $\alpha_{l,nmk}$ are assumed to be symmetric complex Gaussian distributed as $\mathcal{N}\left(0,1\right)$. The power priority weight $\beta_n$ is set to 1, $\forall n$. The target SINRs for all the UEs are the same, i.e., $\gamma_{nk} = \gamma$, $\forall n, k$. Again, the spherical uncertainty model is considered for the CSI errors, i.e., $\mathbf{R}_{mnu} = \epsilon_{mnu}^{-2}\mathbf{I}_{N}$, where $\epsilon_{mnu}^{2} > 0 $ denotes the radius of the uncertainty ball. In order to simulate a realistic asynchronous scenario, we assume that at any CU iteration $i$, the update from each BS arrives with probability $p$, whereas $(1-p)$ denotes the probability that an update is not received at the CU due to either BS failure or network delay. When $p = 1$, the ADBF design reduces to the SDBF design. In our simulation setup, we consider $p=0.6$. Furthermore, we assume that the updates from all the BSs arrive independently at the CU in each iteration.
For the BL-based hybrid TPC design, the AoD space $\boldsymbol{\Phi}_T$ is partitioned into $G = 64$ angular bins. Additionally, the maximum number of EM iterations $\eta_{\mathrm{max}}$ and the stopping parameter $(\rho)$ are set as $50$ and $10^{-5}$, respectively.
\begin{figure*}
\centering
\subfloat[]{\includegraphics[width=0.5\linewidth]{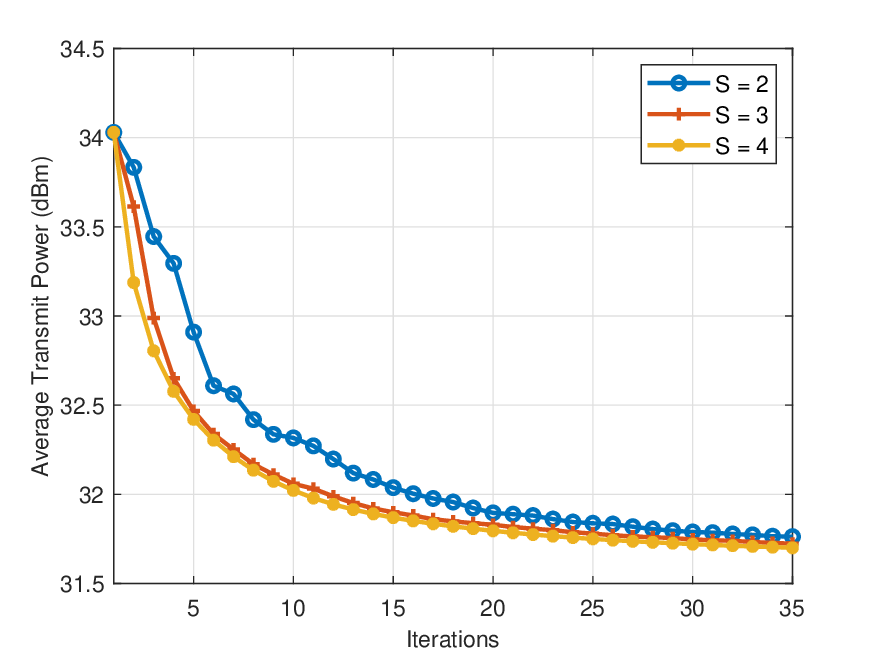}\label{f3a}}
\subfloat[]{\includegraphics[width=0.5\linewidth]{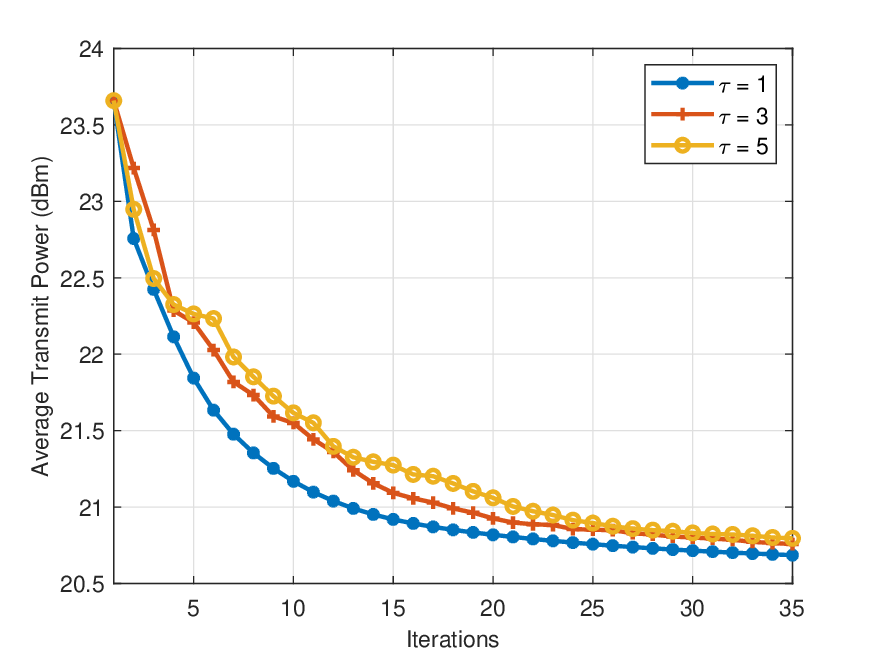}\label{f3b}}
\caption{ Convergence behavior of the ADBF design under various simulation settings: (a) $N=4$, $K=2$, $N_t = 16$, $\epsilon = 0.1$ and $\tau = 4$; (b) $N=4$, $K=2$, $N_t = 16$, $\epsilon = 0.1$ and $S=2$ .}\label{f3}
\end{figure*}
 %Figure 9 starts
\begin{figure*}
\centering
\subfloat[]{\includegraphics[width=0.33\linewidth]{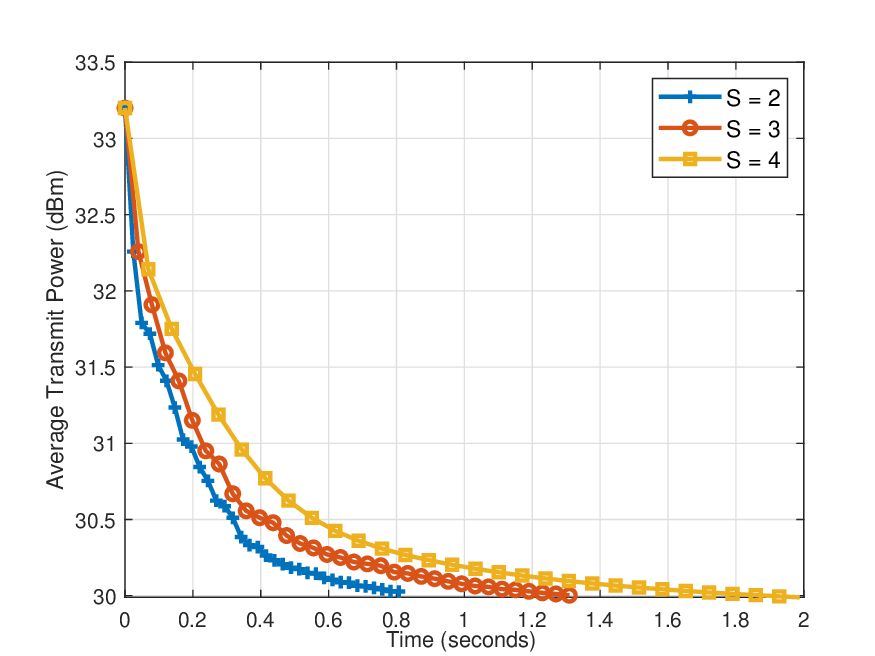}\label{f4a}}
\subfloat[]{\includegraphics[width=0.33\linewidth]{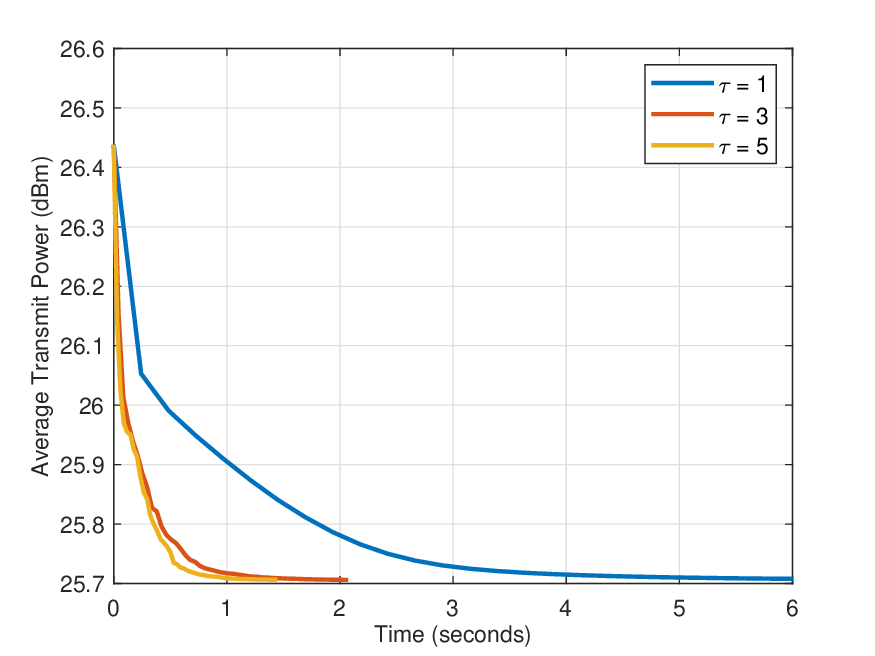}\label{f4b}}
\subfloat[]{\includegraphics[width=0.33\linewidth]{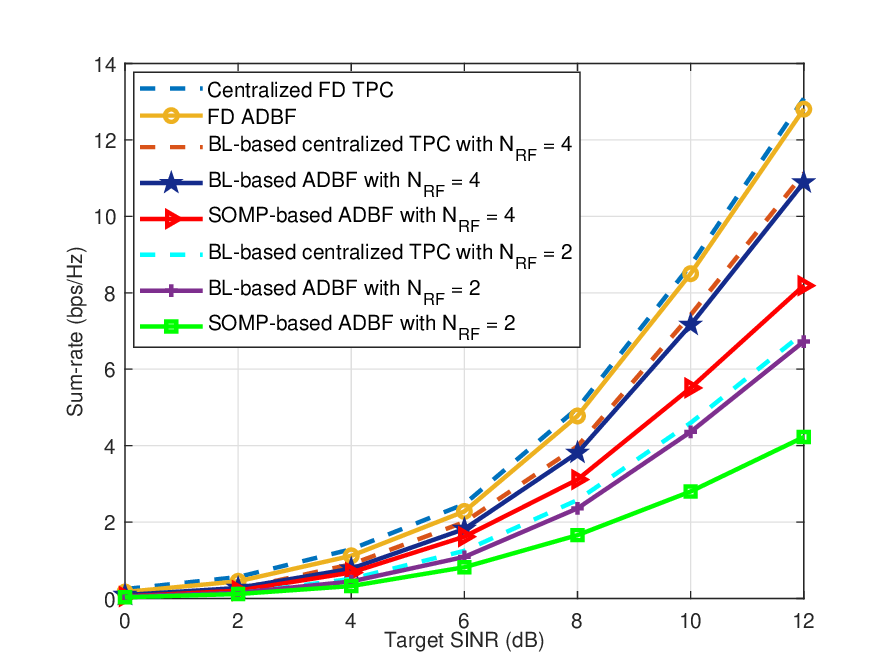}\label{f7a}}
\caption{Convergence behavior of the ADBF design: (a) For different values of $S$ and $\tau=4$; (b) For different values of $\tau$ and $N=4$, $K=2$, $S=2$; (c) Sum rate versus target SINR for $N=3$, $S = 2$, $K=2$, $N_t = 16$ and $\epsilon =0.1$.}\label{f4}
\end{figure*}

Fig. \ref{f2a} compares the feasibility rate of the proposed SDBF and ADBF distributed hybrid beamformers for $N = 2$, $K=2$, $S=1$, $Q=20$ and $\tau = 4$. The feasibility rate is defined as the percentage of the number of successful computations of the corresponding quantities, namely of the average transmit power and beamformer weights using the algorithms proposed in \eqref{eq9}, \eqref{eq20}, \eqref{eq39} and \eqref{eq43}.
One can observe that the ADBF design associated with $S=1$, wherein only one BS update is required to update the global ICI variable $\mathbf{v}$ by the CU, achieves the performance of the SDBF design in which updates from all the BSs are available at the CU simultaneously. Thus, it can be deduced that the proposed ADBF technique is capable of effectively cancelling the ICI with only limited information available at the CU. Due to transmission delays and packet losses that might occur when using the backhaul network for information exchange among the BSs, certain BSs may be using outdated information, forcing the problem to become unfeasible. However, bounded delay conditions and the reliance on only $S$ BS updates used in the algorithm ensure that the updates from all the BSs are incorporated at regular time intervals. This guarantees a high feasibility rate for our ADBF design. Furthermore, the proposed ADBF design also has a significantly enhanced feasibility rate compared to the robust TPC design operating without coordination, which is a non-cooperative beamformer (NCBF) design. This illustrates the superiority of our coordinated ADBF design since the beamformer with no coordination among the BSs fails to cancel the inter-cell interference.
Furthermore, it is interesting to note that the feasibility rate increases upon increasing the number of BS TAs. This arises because the mmWave MCC MIMO system and the ADBF design techniques are able to achieve higher array gain.
Fig. \ref{f1b} plots the feasibility rate versus channel uncertainty radius $\epsilon$ for different target SINR values  $\gamma_{nk}$ of the proposed ADBF design in mmWave MCC networks. As anticipated, it can be observed that the feasibility rate decreases as the CSI uncertainty increases, since it becomes more challenging to meet the minimum SINR requirement when the error radius $\epsilon$ is large. 
Furthermore, it can also be seen that as the desired SINR levels increase, the feasibility rate decreases. This is because the higher SINR demands of the UEs make it challenging to find a feasible solution that satisfies the SINR constraints. Despite this, the proposed ADBF design method still achieves an adequate feasibility rate for high target SINR values even in the face of high CSI uncertainty.

Fig. \ref{f111b} demonstrates the power efficiency of the centralized beamformer and ADBF designs versus the desired target SINR for different number of BS TAs $N_t$ and error radius $\epsilon$. It can be observed that as the target SINR increases, the transmit power must be increased. Nevertheless, as the number of BS TAs increases, the transmit power required decreases due to the enhanced array gain provided by the larger antenna arrays, explicitly highlighting the significance of having a large number of TAs for enhanced power efficiency. Additionally, it is essential to note that the centralized and the ADBF designs utilize the same transmit power. This is because the ADBF design achieves the same performance as the centralized one in significantly less iterations. Finally, it can also be observed that the coordinated ADBF design results in higher power efficiency than the NCBF design at the same target SINR values. Fig. \ref{f4bbaa} once again illustrates the average transmit power (dBm) versus the target SINR requirement for different values of the CSI error of ADBF our design for $N = 4$, $S = 3$ and $K = 3$. Although one can observe that the average transmission power required increases with an increase in the number of cells and users due to the increased ICI, it decreases with an increase in the number of transmit antennas (TAs) at the BSs. This observation emphasizes the importance of deploying a large number of TAs in the mmWave regime to achieve an improved power efficiency.

Fig. \ref{f6a} and  Fig. \ref{f6b} illustrate the sum transmit power of the ADBF design by testing $40$ randomly generated channel realizations. 
The horizontal axes in the figures show the index of the channel realization.
One can observe in \ref{f6a} that $Q = 20$ iterations are adequate for the ADBF algorithm to achieve a performance close to that of the centralized TPC for $N = 2$ cells and the performance does not change further for $Q = 40$ iterations. Upon increasing the number of cells to three, as shown in Fig. \ref{f6b}, $Q = 40$ iterations are seen to be adequate for obtaining a near-optimal performance. This result is similar to that obtained for $80$ iterations. Furthermore, the robust ADBF design achieves a performance close to the perfect CSI scenario.

Let us now define the normalized power accuracy as
 $\frac{\lvert \widehat{P}(i)-P \rvert}{P}$, where $\widehat{P}(i)= \sum_{n=1}^{N}{p_n(i)}$ is the sum transmit power at each iteration $i$ of the distributed beamformer and $P$ denotes the sum transmit power of the centralized TPC. Then Fig. \ref{f5} plots the power accuracy versus the number of iterations required for convergence of the proposed coordinated ADBF design.
 In Fig. \ref{f5a}, $S = 1$ and $S = 2$ correspond to the ADBF and SDBF designs, respectively. Observe that the normalized power accuracy of the ADBF design is close to that of the SDBF design and it achieves an accuracy of $0.01$ in approximately $100$ iterations. Furthermore, Fig. \ref{f5b} and \ref{f5c} compare the performance of the ADBF and SDBF designs for three-cell and four-cell scenarios, respectively. It can be observed that as the value of $S$ increases, i.e., updates from more BSs are incorporated for designing the TPC, the ADBF design closely approaches the performance of the SDBF design. However, the ADBF still achieves a high power accuracy in fewer iterations.

Fig. \ref{f3a} and \ref{f3b} compare the convergence behavior of the ADBF design versus the number of iterations for different values of $S$ and $\tau$, respectively. Observe in Fig. \ref{f3a} that for $S= \{2, 3\}$, the convergence behaviour of the ADBF design closely follows that of the SDBF design ($S = 4$) for a fixed value of $\tau$, where $\tau$ is the bounded delay condition parameter. Explicitly, the update from every BS has to be serviced by the CU at least once in every $\tau$ iterations. This is due to the fact that for a large $S$, local ICI information from more BSs is available at the CU, which leads to convergence in fewer iterations. In Fig. \ref{f3b}, $S$ is fixed at $2$. It can be observed that a large $\tau$ leads to more iterations, since information from the slower BSs is not utilized by the CU frequently. Thus, the number of iterations required for convergence increases upon increasing $\tau$.  
Fig. \ref{f4a} compares the convergence speed of the proposed ADBF and the SDBF designs. It can be observed that the ADBF algorithm converges substantially faster than the SDBF scheme, which is due to the fact that updates from the BSs arrive at the CU more frequently in the former scheme. Similarly, Fig. \ref{f4b} shows the convergence speed of the ADBF algorithm for different values of the parameter $\tau$. It can be observed that as $\tau$ increases, the updates from the slower BSs are not incorporated frequently in each global ICI update by the CU. Therefore, the running time of the ADBF design is shorter than that of its synchronous counterpart ($\tau = 1$). Note that plots in Fig. \ref{f4a} and \ref{f4b} are representative and the time duration for computation in an actual deployment will depend on the computational capabilities of the hardware employed.

Fig. \ref{f7a} compares the sum-rate of the BL-based hybrid TPC design both to that of the ideal FD-TPC and SOMP algorithm \cite{alkhateeb2015limited} based hybrid TPC. One can observe that our BL-based ADBF design provides a significant spectral efficiency gain in comparison to the SOMP-based design. This arises due to the fact that the BL algorithm has improved sparse signal recovery properties in comparison to the SOMP. Furthermore, the performance of the latter scheme is highly sensitive to the choice of the dictionary matrix and stopping criterion. Moreover, one can also observe that the BL-based design attained a performance close to the ideal FD-TPC design, even though it employs significantly fewer RFCs. This can be attributed to the fact that the mmWave MIMO channel has fewer multipath components, which is readily exploited by the proposed BL-based hybrid TPC design. This clearly demonstrates the fact that the ideal FD-TPC can be tightly approximated using a few transmit RFCs.
%\begin{figure}
%\centering
%\includegraphics[width=8cm,height=5cm]{Figure10Latest2.eps}
%\caption{Sum rate versus target SINR for \textcolor{blue}{$N=3$, $S = 2$, $K=2$, $N_t = 16$ and $\epsilon =0.1$}.}\label{f4}
%\end{figure}
\section{Conclusion}\label{conclusions}
Distributed hybrid TPC designs have been proposed for coordinated multi-cell mmWave MIMO systems. Initially, the SDR and BL-based framework has been developed in support of our centralized hybrid TPC design. Next, an ADMM-based asynchronous distributed TPC design was presented, which required only the local CSI and limited information sharing amongst the BSs. Next, robust centralized and ADBF designs were derived by considering realistic scenarios having CSI uncertainty. 
Furthermore, it was also observed that the proposed ADBF design achieves a performance comparable to the centralized solution at a modest signalling overhead, making it ideal for practical implementation. Finally, the simulation results illustrate that the proposed ADBF design is robust against BS failures and network delays.
\bibliographystyle{IEEEtran}
\bibliography{biblio.bib}

% Generated by IEEEtran.bst, version: 1.14 (2015/08/26)
\begin{thebibliography}{10}
\providecommand{\url}[1]{#1}
\csname url@samestyle\endcsname
\providecommand{\newblock}{\relax}
\providecommand{\bibinfo}[2]{#2}
\providecommand{\BIBentrySTDinterwordspacing}{\spaceskip=0pt\relax}
\providecommand{\BIBentryALTinterwordstretchfactor}{4}
\providecommand{\BIBentryALTinterwordspacing}{\spaceskip=\fontdimen2\font plus
\BIBentryALTinterwordstretchfactor\fontdimen3\font minus
  \fontdimen4\font\relax}
\providecommand{\BIBforeignlanguage}[2]{{%
\expandafter\ifx\csname l@#1\endcsname\relax
\typeout{** WARNING: IEEEtran.bst: No hyphenation pattern has been}%
\typeout{** loaded for the language `#1'. Using the pattern for}%
\typeout{** the default language instead.}%
\else
\language=\csname l@#1\endcsname
\fi
#2}}
\providecommand{\BIBdecl}{\relax}
\BIBdecl

\bibitem{shokri2015millimeter}
H.~Shokri-Ghadikolaei, C.~Fischione, G.~Fodor, P.~Popovski, and M.~Zorzi,
  ``Millimeter wave cellular networks: A {MAC} layer perspective,'' \emph{IEEE
  Transactions on Communications}, vol.~63, no.~10, pp. 3437--3458, 2015.

\bibitem{rappaport2015wideband}
T.~S. Rappaport, G.~R. MacCartney, M.~K. Samimi, and S.~Sun, ``Wideband
  millimeter-wave propagation measurements and channel models for future
  wireless communication system design,'' \emph{IEEE Transactions on
  Communications}, vol.~63, no.~9, pp. 3029--3056, 2015.

\bibitem{alkhateeb2015limited}
A.~Alkhateeb, G.~Leus, and R.~W. Heath, ``Limited feedback hybrid precoding for
  multi-user millimeter wave systems,'' \emph{IEEE Transactions on Wireless
  Communications}, vol.~14, no.~11, pp. 6481--6494, 2015.

\bibitem{yong2006overview}
S.~K. Yong and C.-C. Chong, ``An overview of multigigabit wireless through
  millimeter wave technology: Potentials and technical challenges,''
  \emph{EURASIP Journal on Wireless Communications and Networking}, vol. 2007,
  pp. 1--10, 2006.

\bibitem{andrews2016modeling}
J.~G. Andrews, T.~Bai, M.~N. Kulkarni, A.~Alkhateeb, A.~K. Gupta, and R.~W.
  Heath, ``Modeling and analyzing millimeter wave cellular systems,''
  \emph{IEEE Transactions on Communications}, vol.~65, no.~1, pp. 403--430,
  2016.

\bibitem{liang2014low}
L.~Liang, W.~Xu, and X.~Dong, ``Low-complexity hybrid precoding in massive
  multiuser {MIMO} systems,'' \emph{IEEE Wireless Communications Letters},
  vol.~3, no.~6, pp. 653--656, 2014.

\bibitem{boulogeorgos2021coverage}
A.-A.~A. Boulogeorgos and A.~Alexiou, ``Coverage analysis of reconfigurable
  intelligent surface assisted {THz} wireless systems,'' \emph{IEEE Open
  Journal of Vehicular Technology}, vol.~2, pp. 94--110, 2021.

\bibitem{guo2023compressed}
X.~Guo, Y.~Chen, and Y.~Wang, ``Compressed channel estimation for near-field
  {XL-MIMO} using triple parametric decomposition,'' \emph{IEEE Transactions on
  Vehicular Technology}, 2023.

\bibitem{satyanarayana2018hybrid}
K.~Satyanarayana, M.~El-Hajjar, P.-H. Kuo, A.~Mourad, and L.~Hanzo, ``Hybrid
  beamforming design for full-duplex millimeter wave communication,''
  \emph{IEEE Transactions on Vehicular Technology}, vol.~68, no.~2, pp.
  1394--1404, 2018.

\bibitem{pi2011introduction}
Z.~Pi and F.~Khan, ``An introduction to millimeter-wave mobile broadband
  systems,'' \emph{IEEE Communications Magazine}, vol.~49, no.~6, pp. 101--107,
  2011.

\bibitem{sohrabi2016hybrid}
F.~Sohrabi and W.~Yu, ``Hybrid digital and analog beamforming design for
  large-scale antenna arrays,'' \emph{IEEE Journal of Selected Topics in Signal
  Processing}, vol.~10, no.~3, pp. 501--513, 2016.

\bibitem{shafi20175g}
M.~Shafi, A.~F. Molisch, P.~J. Smith, T.~Haustein, P.~Zhu, P.~De~Silva,
  F.~Tufvesson, A.~Benjebbour, and G.~Wunder, ``{5G}: A tutorial overview of
  standards, trials, challenges, deployment, and practice,'' \emph{IEEE Journal
  on Selected Areas in Communications}, vol.~35, no.~6, pp. 1201--1221, 2017.

\bibitem{singh2023energy}
J.~Singh, S.~Srivastava, S.~P. Yadav, A.~K. Jagannatham, and L.~Hanzo, ``Energy
  efficiency optimization in reconfigurable intelligent surface aided hybrid
  multiuser {mmWave} {MIMO} systems,'' \emph{IEEE Open Journal of Vehicular
  Technology}, 2023.

\bibitem{bai2014coverage}
T.~Bai and R.~W. Heath, ``Coverage and rate analysis for millimeter-wave
  cellular networks,'' \emph{IEEE Transactions on Wireless Communications},
  vol.~14, no.~2, pp. 1100--1114, 2014.

\bibitem{murdock2013consumption}
J.~N. Murdock and T.~S. Rappaport, ``Consumption factor and power-efficiency
  factor: A theory for evaluating the energy efficiency of cascaded
  communication systems,'' \emph{IEEE Journal on Selected Areas in
  Communications}, vol.~32, no.~2, pp. 221--236, 2013.

\bibitem{jafar2004phantomnet}
S.~A. Jafar, G.~J. Foschini, and A.~J. Goldsmith, ``Phantomnet: Exploring
  optimal multicellular multiple antenna systems,'' \emph{EURASIP Journal on
  Advances in Signal Processing}, vol. 2004, no.~5, pp. 1--14, 2004.

\bibitem{marsch2008multicell}
P.~Marsch and G.~Fettweis, ``On multicell cooperative transmission in
  backhaul-constrained cellular systems,'' \emph{Annals of
  telecommunications-annales des t{\'e}l{\'e}communications}, vol.~63, no.~5,
  pp. 253--269, 2008.

\bibitem{chrisT}
C.~T.~K. Ng and H.~Huang, ``Linear precoding in cooperative {MIMO} cellular
  networks with limited coordination clusters,'' \emph{IEEE Journal on Selected
  Areas in Communications}, vol.~28, no.~9, pp. 1446--1454, 2010.

\bibitem{zhang2010cooperative}
R.~Zhang and L.~Hanzo, ``Cooperative downlink multicell preprocessing relying
  on reduced-rate back-haul data exchange,'' \emph{IEEE Transactions on
  Vehicular Technology}, vol.~60, no.~2, pp. 539--545, 2010.

\bibitem{xiang2012coordinated}
Z.~Xiang, M.~Tao, and X.~Wang, ``Coordinated multicast beamforming in multicell
  networks,'' \emph{IEEE Transactions on Wireless Communications}, vol.~12,
  no.~1, pp. 12--21, 2012.

\bibitem{otter}
S.~He, Y.~Huang, L.~Yang, and B.~Ottersten, ``Coordinated multicell multiuser
  precoding for maximizing weighted sum energy efficiency,'' \emph{IEEE
  Transactions on Signal Processing}, vol.~62, no.~3, pp. 741--751, 2014.

\bibitem{lakshminarayana2015coordinated}
S.~Lakshminarayana, M.~Assaad, and M.~Debbah, ``Coordinated multicell
  beamforming for massive {MIMO}: A random matrix approach,'' \emph{IEEE
  Transactions on Information Theory}, vol.~61, no.~6, pp. 3387--3412, 2015.

\bibitem{xie2017robust}
X.~Xie, H.~Yang, and A.~V. Vasilakos, ``Robust transceiver design based on
  interference alignment for multi-user multi-cell {MIMO} networks with channel
  uncertainty,'' \emph{IEEE Access}, vol.~5, pp. 5121--5134, 2017.

\bibitem{CRAN}
O.~Dhif-Allah, H.~Dahrouj, T.~Y. Al-Naffouri, and M.-S. Alouini, ``Distributed
  robust power minimization for the downlink of multi-cloud radio access
  networks,'' \emph{IEEE Transactions on Green Communications and Networking},
  vol.~2, no.~2, pp. 327--335, 2018.

\bibitem{chen2020multiobjective}
W.-Y. Chen, B.-S. Chen, and W.-T. Chen, ``Multiobjective beamforming power
  control for robust {SINR} target tracking and power efficiency in multicell
  {MU}-{MIMO} wireless system,'' \emph{IEEE Transactions on Vehicular
  Technology}, vol.~69, no.~6, pp. 6200--6214, 2020.

\bibitem{chen2021robust}
Y.~Chen, Y.~Wang, and L.~Jiao, ``Robust transmission for reconfigurable
  intelligent surface aided millimeter wave vehicular communications with
  statistical {CSI},'' \emph{IEEE Transactions on Wireless Communications},
  vol.~21, no.~2, pp. 928--944, 2021.

\bibitem{chen2022reconfigurable}
Y.~Chen, Y.~Wang, and Z.~Wang, ``Reconfigurable intelligent surface aided
  high-mobility millimeter wave communications with dynamic dual-structured
  sparsity,'' \emph{IEEE Transactions on Wireless Communications}, 2022.

\bibitem{xu2017spectral}
W.~Xu, J.~Liu, S.~Jin, and X.~Dong, ``Spectral and energy efficiency of
  multi-pair massive {MIMO} relay network with hybrid processing,'' \emph{IEEE
  Transactions on Communications}, vol.~65, no.~9, pp. 3794--3809, 2017.

\bibitem{michaloliakos2016joint}
A.~Michaloliakos, W.-C. Ao, and K.~Psounis, ``Joint user-beam selection for
  hybrid beamforming in asynchronously coordinated multi-cell networks,'' in
  \emph{2016 Information Theory and Applications Workshop (ITA)}.\hskip 1em
  plus 0.5em minus 0.4em\relax IEEE, 2016, pp. 1--10.

\bibitem{sun2018analytical}
S.~Sun, T.~S. Rappaport, M.~Shafi, and H.~Tataria, ``Analytical framework of
  hybrid beamforming in multi-cell millimeter-wave systems,'' \emph{IEEE
  Transactions on Wireless Communications}, vol.~17, no.~11, pp. 7528--7543,
  2018.

\bibitem{castanheira2017hybrid}
D.~Castanheira, P.~Lopes, A.~Silva, and A.~Gameiro, ``Hybrid beamforming
  designs for massive {MIMO} millimeter-wave heterogeneous systems,''
  \emph{IEEE Access}, vol.~5, pp. 21\,806--21\,817, 2017.

\bibitem{kumar2021blockage}
D.~Kumar, J.~Kaleva, and A.~T{\"o}lli, ``Blockage-aware reliable mm{W}ave
  access via coordinated multi-point connectivity,'' \emph{IEEE Transactions on
  Wireless Communications}, vol.~20, no.~7, pp. 4238--4252, 2021.

\bibitem{bai2018cooperative}
L.~Bai, T.~Li, Q.~Yu, J.~Choi, and W.~Zhang, ``Cooperative multiuser
  beamforming in mm{W}ave distributed antenna systems,'' \emph{IEEE
  Transactions on Vehicular Technology}, vol.~67, no.~12, pp. 12\,394--12\,397,
  2018.

\bibitem{zhao2020robust}
M.-M. Zhao, Y.~Cai, M.-J. Zhao, Y.~Xu, and L.~Hanzo, ``Robust joint hybrid
  analog-digital transceiver design for full-duplex mm{W}ave multicell
  systems,'' \emph{IEEE Transactions on Communications}, vol.~68, no.~8, pp.
  4788--4802, 2020.

\bibitem{jafri2022robust}
M.~Jafri, A.~Anand, S.~Srivastava, A.~K. Jagannatham, and L.~Hanzo, ``Robust
  distributed hybrid beamforming in coordinated multi-user multi-cell mm{W}ave
  {MIMO} systems relying on imperfect {CSI},'' \emph{IEEE Transactions on
  Communications}, vol.~70, no.~12, pp. 8123--8137, 2022.

\bibitem{jafri2023cooperative}
M.~Jafri, S.~Srivastava, N.~K. Venkategowda, A.~K. Jagannatham, and L.~Hanzo,
  ``Cooperative hybrid transmit beamforming in cell-free mm{W}ave {MIMO}
  networks,'' \emph{IEEE Transactions on Vehicular Technology}, 2023.

\bibitem{sha2022near}
Z.~Sha, S.~Chen, and Z.~Wang, ``Near interference-free space-time user
  scheduling for mmwave cellular network,'' \emph{IEEE Transactions on Wireless
  Communications}, vol.~21, no.~8, pp. 6372--6386, 2022.

\bibitem{sha2023versatile}
Z.~Sha, Y.~Ming, C.~Sun, and Z.~Wang, ``Versatile resource management for
  millimeter-wave cellular network: Near interference-free scheduling
  methodology,'' \emph{IEEE Transactions on Vehicular Technology}, 2023.

\bibitem{wipf2004sparse}
D.~P. Wipf and B.~D. Rao, ``Sparse {B}ayesian learning for basis selection,''
  \emph{IEEE Transactions on Signal Processing}, vol.~52, no.~8, pp.
  2153--2164, 2004.

\bibitem{boyd2004convex}
S.~Boyd, S.~P. Boyd, and L.~Vandenberghe, \emph{Convex optimization}.\hskip 1em
  plus 0.5em minus 0.4em\relax Cambridge University Press, 2004.

\bibitem{luo2010sdp}
Z.-Q. Luo, T.-H. Chang, D.~Palomar, and Y.~Eldar, ``{SDP} relaxation of
  homogeneous quadratic optimization: approximation,'' \emph{Convex
  Optimization in Signal Processing and Communications}, p. 117, 2010.

\bibitem{grant2014cvx}
M.~Grant and S.~Boyd, ``{CVX}: Matlab software for disciplined convex
  programming, version 2.1,'' 2014.

\bibitem{heath2016overview}
R.~W. Heath, N.~Gonzalez-Prelcic, S.~Rangan, W.~Roh, and A.~M. Sayeed, ``An
  overview of signal processing techniques for millimeter wave {MIMO}
  systems,'' \emph{{IEEE} Journal of Selected Topics in Signal Processing},
  vol.~10, no.~3, pp. 436--453, 2016.

\bibitem{hestenes1969multiplier}
M.~R. Hestenes, ``Multiplier and gradient methods,'' \emph{Journal of
  Optimization Theory and Applications}, vol.~4, no.~5, pp. 303--320, 1969.

\bibitem{boyd2011distributed}
S.~Boyd, N.~Parikh, and E.~Chu, \emph{Distributed optimization and statistical
  learning via the alternating direction method of multipliers}.\hskip 1em plus
  0.5em minus 0.4em\relax Now Publishers Inc, 2011.

\bibitem{zhang2011unified}
X.~Zhang, M.~Burger, and S.~Osher, ``A unified primal-dual algorithm framework
  based on {Bregman} iteration,'' \emph{Journal of Scientific Computing},
  vol.~46, pp. 20--46, 2011.

\bibitem{yoon1988lower}
S.~Yoon and A.~Jameson, ``Lower-upper symmetric-gauss-seidel method for the
  euler and navier-stokes equations,'' \emph{{AIAA} journal}, vol.~26, no.~9,
  pp. 1025--1026, 1988.

\bibitem{converg}
W.~Jiang, A.~Grammenos, E.~Kalyvianaki, and T.~Charalambous, ``An asynchronous
  approximate distributed alternating direction method of multipliers in
  digraphs,'' in \emph{2021 60th IEEE Conference on Decision and Control
  (CDC)}, 2021, pp. 3406--3413.

\bibitem{async}
J.~Zhou and Y.~Lei, ``Asynchronous group-based {ADMM} algorithm under efficient
  communication structure,'' in \emph{2018 IEEE Intl Conf on Parallel {\&}
  Distributed Processing with Applications, Ubiquitous Computing {\&}
  Communications, Big Data {\&} Cloud Computing, Social Computing {\&}
  Networking, Sustainable Computing {\&} Communications
  (ISPA/IUCC/BDCloud/SocialCom/SustainCom)}, 2018, pp. 135--140.

\end{thebibliography}
\end{document}